\documentclass[journal,twoside]{IEEEtran}
\usepackage{amsthm}
\usepackage{amsmath,amsfonts}
\usepackage{algorithmic}
\usepackage{algorithm}
\usepackage{array}
\usepackage{booktabs}
\usepackage{bm}
\usepackage[caption=false,font=footnotesize,labelfont=rm,textfont=rm]{subfig}
\usepackage{color}
\usepackage{mathtools}
\usepackage{setspace}
\usepackage{textcomp}
\usepackage{tcolorbox}
\usepackage{threeparttable}
\usepackage{makecell}
\usepackage{multirow}
\usepackage{stfloats}
\usepackage{url}
\usepackage{verbatim}
\usepackage{graphicx}
\usepackage{cite}
\usepackage{epsfig}
\usepackage{epstopdf}

\newcommand{\T}{\mathrm{T}}

\begin{document}
\setstretch{0.905}

\title{Grid-following and Grid-forming Switching Control for Grid-connected Inverters Considering Small-signal Security Region \vspace{-2mm}}

\author{
	Qiping Lai,~\IEEEmembership{Graduate Student Member,~IEEE}, Yi Shen,~\IEEEmembership{Graduate Student Member,~IEEE} \\ and Chen Shen,~\IEEEmembership{Senior Member,~IEEE}
	\thanks{This work was supported by Smart Grid-National Science and Technology Major Project (2024ZD0801500). \textit{(Corresponding author: Chen Shen.)}}
	\thanks{The authors are with the Department of Electrical Engineering, Tsinghua University, Beijing 100084, China (e-mail: lqp22@mails.tsinghua.edu.cn; yi-shen23@mails.tsinghua.edu.cn; shenchen@mail.tsinghua.edu.cn).}
    \vspace{-4mm}
}


\maketitle
\begin{abstract}
In high-penetration renewable power systems with complex and highly variable operating scenarios, grid-connected inverters (GCIs) may transition between different control modes to adapt to diverse grid conditions. Among these, the switching between grid-following (GFL) and grid-forming (GFM) control modes is particularly critical. Nevertheless, safe and robust GFL-GFM switching control strategies for GCIs remain largely unexplored. To overcome this challenge, this paper establishes a full-order small-signal state-space model for the GFL-GFM switched system, precisely reflecting all internal circuit and control dynamics. Subsequently, the small-signal security region (SSSR) of the switched system is defined and characterized, followed by an in-depth investigation into the multi-parameter impacts on the SSSRs and internal stability margin distributions (ISMDs). Furthermore, a novel comprehensive stability index (CSI) is proposed by integrating the stability margin, parameter sensitivity, and boundary distance. Based on this CSI, a multi-objective adaptive GFL-GFM switching control strategy is designed to guarantee the dynamic security and robustness of the system. Finally, the proposed SSSR analysis method for the GFL-GFM switched system and the designed CSI-based switching control mechanism are validated through electromagnetic transient (EMT) simulations.
\end{abstract}


\section{Introduction}
\IEEEPARstart{M}{odern} power systems are rapidly evolving toward a new paradigm characterized by high penetrations of renewable energy sources and the extensive deployment of power electronic devices \cite{Chen_guesteditorialmodels_2023,Li_mianxiangtan_2021}. While the massive integration of inverter-based resources significantly enhances the flexibility and economic efficiency of the grid, it profoundly alters the dynamic characteristics of the power system \cite{Sharma_parametererroridentification_2026,Xie_shuanggaodian_2021}. As a consequence, the grid exhibits heightened uncertainty and complexity, leading to the frequent emergence of small-signal stability issues in grid-connected inverter (GCI) systems \cite{Pan_robustsmallsignalstability_2018}.

Due to the fundamentally different synchronization mechanisms, grid-following (GFL) and grid-forming (GFM) inverters present complementary stability performance under varying short circuit ratios (SCRs) \cite{Rosso_gridformingconverterscontrol_2021,LiuHui_gouwangxing_2025}. Specifically, GFL inverters track the voltage phase at the point of common coupling through a phase-locked loop (PLL), demonstrating robust stability in strong grids but becoming highly susceptible to oscillatory instability under weak grid conditions \cite{Zhao_voltagedynamicscurrent_2016}. In contrast, GFM inverters employ power synchronization control to autonomously establish their internal voltage magnitude and phase. This mechanism ensures superior stability in weak grids but leads to low stability margins in strong grids \cite{Poolla_placementimplementationgridforming_2019}. Therefore, leveraging their complementary strengths through GFL-GFM hybrid control has become a critical research focus for the safe and stable operation of modern power systems \cite{ZhangXing_xinnengyuan_2021}.

Recent studies have proposed a variety of GFL-GFM hybrid control strategies, mainly including switching-based \cite{Li_impedanceadaptivedualmode_2021}, integrated \cite{Lima_hybridcontrolscheme_2022}, compensation-based \cite{Moutevelis_recursivesecondarycontroller_2023}, synchronous \cite{Meng_generalizeddroopcontrol_2019}, and plant-level \cite{Tarraso_designcontrollervirtual_2021} hybrid control schemes. Under the GFL-GFM switching control, the GCI adopts the GFL mode under strong grid conditions and switches to the GFM mode in weak grids. Owing to its direct implementation and robust adaptability in utilizing complementary advantages of both control modes, this switching control approach has attracted widespread research interest \cite{Zhang_gaoshentou_2024}. In \cite{Ding_noveldesignswitchable_2025}, a switchable GFL-GFM control architecture for voltage source inverters is developed and experimentally validated on a hardware testbed, but it lacks an adaptive switching mechanism. As indicated in \cite{Li_gaoshentou_2021}, to achieve an undisturbed transition between GFL and GFM operations, both the phase angles and the inner current loop references of two control modes must be identical at the switching instant. Reference \cite{Song_improvedmixtureratio_2025} proposes an improved mixture ratio control strategy for zero-disturbance switching of GFL and GFM inverters, enabling both control loops to operate in a closed loop at the same operating point by calculating the virtual current and virtual voltage. Nevertheless, the explicit switching boundary remains unaddressed in this work. In \cite{Li_impedanceadaptivedualmode_2021,Li_gaoshentou_2021}, the D-partition method \cite{Hwang_robustdpartition_2010} is employed to establish the parameter stability regions and switching boundary for GFL and GFM control modes under multiple performance constraints, including stability margins and control loop bandwidths. Although this impedance adaptive dual-mode control approach improves robustness against large SCR fluctuations, its linear transfer function basis limits the number of considered parameters and
state variables \cite{Jinggong_calculationpicontroller_2010}. Considering the impacts of grid impedance and power output on system stability, reference \cite{ZhouYuQing_jiyuyun_2024} proposes an optimal configuration method for GFL-GFM switchable units based on the operating SCR. While this cost-effective scheme can be utilized for retrofitting existing renewable plants to meet stability demands, it fails to incorporate the inverter circuit dynamics and internal control parameters.

In summary, GFL-GFM switching control serves as an effective grid-adaptive control paradigm that switches the overall control structure, allowing for the direct utilization of existing GFL and GFM control strategies to select the optimal operating mode according to distinct system conditions \cite{LiMing_gaoshentou_2024}. Most existing studies rely on reduced-order inverter models to investigate switching boundaries, primarily adopting the SCR as the switching indicator \cite{Wang_powerselfsynchronizationcontrol_2023}. However, there is still a lack of quantitative analysis regarding the stability and robustness of GCIs under the compounded influence of multiple parameters. Therefore, the design of robust switching control strategies considering multiple parameters and indicators remains an open challenge. To bridge this gap, this paper proposes a novel multi-objective adaptive GFL-GFM switching control strategy for GCIs, which is systematically designed based on a comprehensive stability index (CSI) derived from the small-signal security region (SSSR) and its internal stability margin distribution (ISMD). The main contributions of this paper are summarized as follows:
\begin{enumerate}
	\item {A full-order small-signal state-space model for GCIs under GFL and GFM switching control is constructed, systematically capturing the coupled impacts of hardware component parameters, inverter control parameters, and grid connection impedance on system stability. This detailed switched system model provides a solid theoretical foundation for the advanced parameter formulation and control design of GCIs.}
	\item {The SSSR for the GFL-GFM switched system is defined and geometrically characterized via a hyperplane-approximation approach. By analyzing the influences of multiple system parameters on the SSSR boundaries and ISMDs, this work enables optimal parameter configuration to enhance the system's small-signal stability.}
	\item {To guarantee the dynamic security and robustness of GCIs under varying grid conditions, a multi-objective adaptive GFL-GFM switching control strategy driven by a novel CSI is proposed. The CSI is formulated by integrating multiple normalized indicators, including the stability margin, parameter sensitivity, and boundary distance, all of which are analytically derived using a Gaussian mixture model (GMM).}
\end{enumerate}

The remainder of this paper is organized as follows. Section II constructs the full-order switched system model of GCIs under GFL and GFM switching control. Section III characterizes the SSSR and ISMD for GFL and GFM inverters. Section IV proposes a novel CSI-based switching control strategy for GFL-GFM switched system. Section V verifies the proposed switched system model, SSSR analysis approach and switching strategy through electromagnetic transient (EMT) simulation. Finally, conclusions are provided in Section VI.

\section{Switched System Model of the GCIs Under GFL and GFM Switching Control}
\begin{figure*}[!t]
    \centering
    \vspace{-3mm}
    \includegraphics[width=1.4\columnwidth]{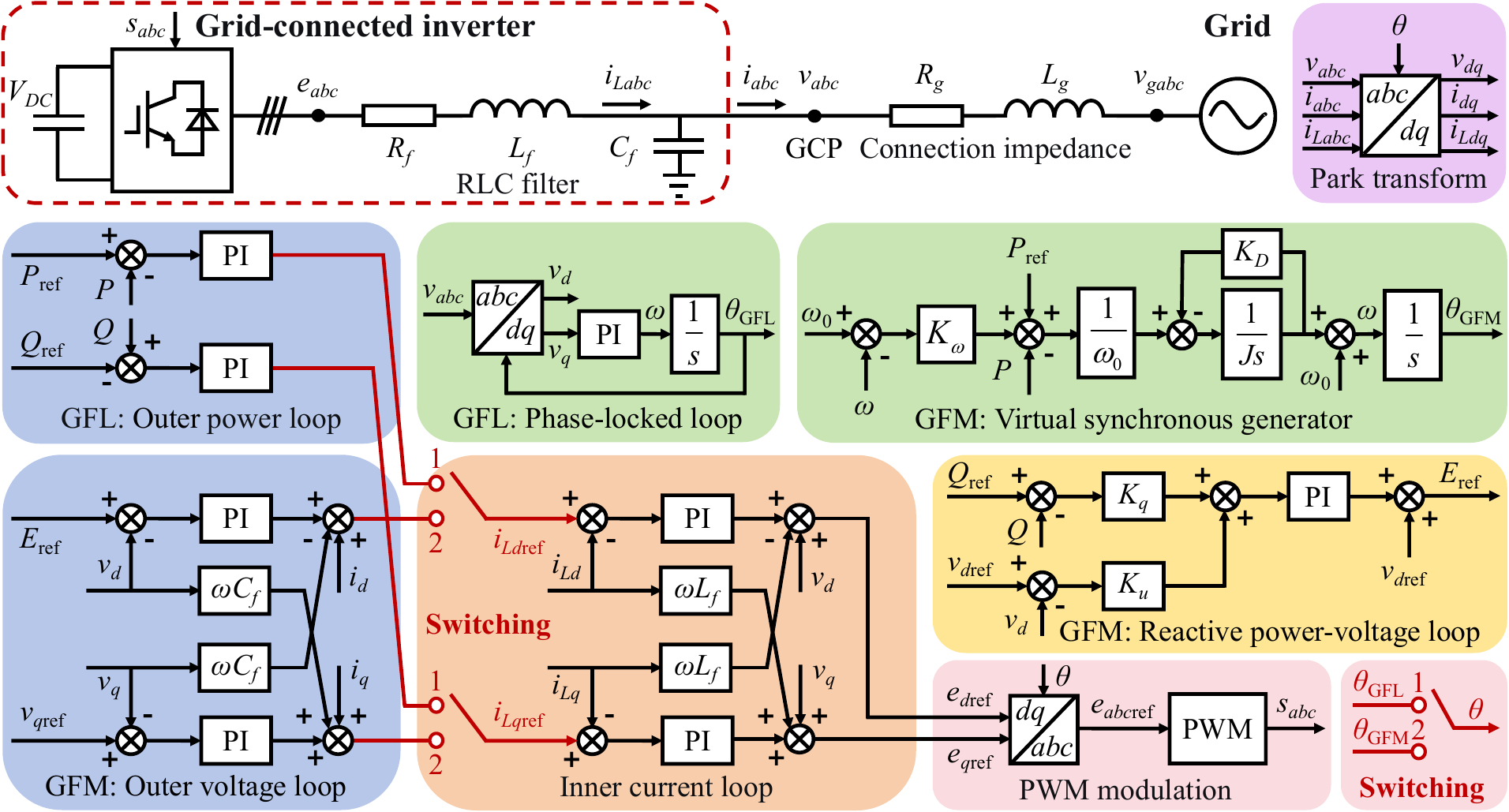}
    \vspace{-2mm}
    \caption{The overall architecture diagram of GCIs under GFL and GFM switching control.}
    \vspace{-2mm}
    \label{Fig1_GFLGFM_Switched_system}
\end{figure*}

The overall architecture of GCIs operating under GFL and GFM switching control is depicted in Fig. \ref{Fig1_GFLGFM_Switched_system} \cite{QuZiSen_gaobili_2021,Gao_seamlessswitchingmethod_2025}. This section first presents the per-unit mathematical model for each fundamental module based on this architecture. Subsequently, small-signal linearization is applied to derive the full-order state-space representations of GFL and GFM control subsystems. Finally, by incorporating a defined switching law to orchestrate the mode transitions, the GFL-GFM switched system model for the GCI is established. 

\subsection{Mathematical Model of the Electrical Circuits}
As illustrated in Fig. \ref{Fig1_GFLGFM_Switched_system}, various grid-connected power electronic devices, such as wind turbines, photovoltaic systems, energy storage units, and static var generators, share a fundamental alternating current (AC) electrical topology. Their direct current (DC) power is interfaced with the AC grid via voltage source inverters (VSIs). The AC output is delivered to the main grid through the filter and line connection impedance.

The mathematical model describing the inverter's AC-side filter dynamics in the $dq$ rotating reference frame can be expressed as
\begin{equation} \label{eqn1}
    \begin{cases}
    \dfrac{{{\rm{d}}{i_{Ld}}}}{{{\rm{d}}t}} = \dfrac{1}{{{L_f}}}\left( {{e_d} - {v_d} + \omega {L_f}{i_{Lq}} - {R_f}{i_{Ld}}} \right)\\[2mm]
    \dfrac{{{\rm{d}}{i_{Lq}}}}{{{\rm{d}}t}} = \dfrac{1}{{{L_f}}}\left( {{e_q} - {v_q} - \omega {L_f}{i_{Ld}} - {R_f}{i_{Lq}}} \right)\\[2mm]
    \dfrac{{{\rm{d}}{v_d}}}{{{\rm{d}}t}} = \dfrac{1}{{{C_f}}}\left( {{i_{Ld}} - {i_d} + \omega {C_f}{v_q}} \right)\\[2mm]
    \dfrac{{{\rm{d}}{v_q}}}{{{\rm{d}}t}} = \dfrac{1}{{{C_f}}}\left( {{i_{Lq}} - {i_q} - \omega {C_f}{v_d}} \right)
    \end{cases}
\end{equation}
where $\omega$ is the inverter's electrical angular frequency; $R_f$, $L_f$, and $C_f$ denote the filter resistance, inductance, and capacitance, respectively; $e_d$ and $e_q$ represent the $d$- and $q$-axis components of the inverter's AC-side voltage, respectively; $i_{Ld}$ and $i_{Lq}$ are the $d$- and $q$-axis components of the inverter's AC-side current, respectively; $v_d$ and $v_q$ indicate the $d$- and $q$-axis components of the grid connection point (GCP) voltage, respectively; $i_d$ and $i_q$ correspond to the $d$- and $q$-axis components of the grid-side injected current, respectively.

The mathematical model of the grid connection impedance in the $dq$ rotating reference frame is formulated as
\begin{equation} \label{eqn2}
    \begin{cases}
    {v_{gd}} = {v_g}\cos \delta, \quad {v_{gq}} = - {v_g}\sin \delta \\[2mm]
    \dfrac{{{\rm{d}}{i_d}}}{{{\rm{d}}t}} = \dfrac{1}{{{L_g}}}\left( {{v_d} - {v_{gd}} + \omega {L_g}{i_q} - {R_g}{i_d}} \right)\\[2mm]
    \dfrac{{{\rm{d}}{i_q}}}{{{\rm{d}}t}} = \dfrac{1}{{{L_g}}}\left( {{v_q} - {v_{gq}} - \omega {L_g}{i_d} - {R_g}{i_q}} \right)
    \end{cases}
\end{equation}
where $v_g$ represents the amplitude of the grid voltage vector; $\delta$ denotes the phase angle that $d$-axis leads the grid voltage vector; $v_{gd}$ and $v_{gq}$ indicate the $d$- and $q$-axis components of the grid voltage, respectively; $R_g$ and $L_g$ are the equivalent line resistance and inductance, respectively.

\subsection{Mathematical Model of the Control Modules}
As shown in Fig. \ref{Fig1_GFLGFM_Switched_system}, GFL and GFM inverters adopt identical coordinate transformation, power calculation, and modulation modules. However, they differ fundamentally in synchronization mechanisms and dual-loop control strategies.

\subsubsection{Common Control Components}
Through the Park transformation, the three-phase electrical quantities are converted into the two-phase rotating reference frame to facilitate the decoupled vector control of the inverter's voltage and current. The $abc-dq0$ transformation matrix is presented as
\begin{equation} \label{eqn3}
    T = \dfrac{2}{3}\begin{bmatrix}
    {\cos \theta}&{\cos (\theta - 2\pi /3)}&{\cos (\theta  + 2\pi /3)}\\
    { - \sin \theta }&{ - \sin (\theta  - 2\pi /3)}&{ - \sin (\theta  + 2\pi /3)}\\
    {1/2}&{1/2}&{1/2}
    \end{bmatrix}
\end{equation}
where $\theta$ represents the phase angle between the $a$-axis of the $abc$ stationary frame and the $d$-axis of the $dq$ rotating frame.

Moreover, the active and reactive power in the per-unit (p.u.) system are calculated as
\begin{equation} \label{eqn4}
    \begin{cases}
    {P = {v_d}{i_d} + {v_q}{i_q}}\\
    {Q = {v_q}{i_d} - {v_d}{i_q}}
    \end{cases}
\end{equation}
where $P$ and $Q$ denote the active and reactive power outputs of the inverter, respectively.

In the modulation block, the inverter's AC-side voltage references in the $dq$ frame (i.e., $e_{d\mathrm{ref}}$, $e_{q\mathrm{ref}}$) are transformed back to the $abc$ frame using the inverse Park transformation. These voltage references are assumed to be perfectly tracked by the actual $d$- and $q$-axis voltages due to the high response frequency of the modulation block with wide bandwidth \cite{LiuHui_gouwangxing_2025}. 


\subsubsection{GFL Control Loops}
As depicted in Fig. \ref{Fig1_GFLGFM_Switched_system}, the GFL control scheme primarily comprises a phase-locked loop (PLL), an outer power loop (OPL), and an inner current loop (ICL).

The PLL employs the grid-voltage-oriented vector control technique, which aligns the $d$-axis of the $dq$ frame with the GCP voltage vector to ensure precise phase synchronization. 
The mathematical model of the PLL is expressed as
\begin{equation} \label{eqn5}
    \begin{cases}
        \dfrac{{{\rm{d}}\zeta }}{{{\rm{d}}t}} = {v_q}, \quad \dfrac{{{\rm{d}}\delta }}{{{\rm{d}}t}} = \omega  - {\omega _0}\\
        \omega  = {K_{p\rm{PLL}}}{v_q} + {K_{i\rm{PLL}}} \zeta
    \end{cases}
\end{equation}
where $\zeta$ represents the integral state variable of the PI integrator; $\omega_0$ denotes the synchronous electrical angular frequency of the grid; ${K_{p\rm{PLL}}}$ and ${K_{i\rm{PLL}}}$ are the proportional and integral gains of the PLL, respectively.

The OPL of the GFL inverter adopts constant power control. The deviations between the measured active/reactive powers and their respective setpoints are processed by PI controllers to generate the $d$- and $q$-axis current references. Then, the ICL tracks these references to produce the inverter's AC-side voltage references. Notably, feedforward decoupling terms are integrated into the ICL to eliminate the cross-coupling dynamics of the filter. The mathematical model of this power-current dual-loop control scheme is derived as
\begin{equation} \label{eqn6}
    \begin{cases}
        \dfrac{{{\rm{d}}{\gamma _d}}}{{{\rm{d}}t}} = {P_{{\rm{ref}}}} - P, \quad \dfrac{{{\rm{d}}{\gamma _q}}}{{{\rm{d}}t}} = Q - {Q_{{\rm{ref}}}}\\[2mm]
        {i_{Ld{\rm{ref}}}} = {K_{po1}}\left( {{P_{{\rm{ref}}}} - P} \right) + {K_{io1}}{\gamma _d}\\[2mm]
        {i_{Lq{\rm{ref}}}} = {K_{po1}}\left( {Q - {Q_{{\rm{ref}}}}} \right) + {K_{io1}}{\gamma _q}\\[2mm]        
        \dfrac{{{\rm{d}}{\xi _d}}}{{{\rm{d}}t}} = {i_{Ld{\rm{ref}}}} - {i_{Ld}}, \quad \dfrac{{{\rm{d}}{\xi _q}}}{{{\rm{d}}t}} = {i_{Lq{\rm{ref}}}} - {i_{Lq}}\\[2mm]
        {e_{d{\rm{ref}}}} = {v_d} - \omega {L_f}{i_{Lq}} + {K_{pi1}}\left( {{i_{Ld{\rm{ref}}}} - {i_{Ld}}} \right) + {K_{ii1}}{\xi _d}\\[2mm]
        {e_{q{\rm{ref}}}} = {v_q} + \omega {L_f}{i_{Ld}} + {K_{pi1}}\left( {{i_{Lq{\rm{ref}}}} - {i_{Lq}}} \right) + {K_{ii1}}{\xi _q}
    \end{cases}\!\!\!\!\!\!\!\!\!\!
\end{equation}
where $\gamma_d$, $\gamma_q$, $\xi_d$ and $\xi_q$ are all integral state variables; $P_{\mathrm{ref}}$ and $Q_{\mathrm{ref}}$ denote the active and reactive power references, respectively; $i_{Ld\mathrm{ref}}$ and $i_{Lq\mathrm{ref}}$ represent the $d$- and $q$-axis components of the inverter's AC-side current references, respectively; ${K_{po1}}$ and ${K_{io1}}$ indicate the proportional and integral gains of the OPL, respectively; ${K_{pi1}}$ and ${K_{ii1}}$ correspond to the proportional and integral gains of the ICL, respectively.

\subsubsection{GFM Control Loops}
As depicted in Fig. \ref{Fig1_GFLGFM_Switched_system}, the GFM control scheme mainly consists of a power synchronization loop (exemplified by virtual synchronous generator (VSG) control), an outer voltage loop (OVL), and an ICL.

VSG control emulates the electromechanical characteristics of conventional synchronous generators (SGs), enabling GFM inverters to actively participate in grid frequency regulation, voltage adjustment, and inertia support. Its fundamental mechanism relies on the incorporation of virtual inertia, virtual damping, and droop characteristics, allowing the inverter to exhibit SG-like dynamic behaviors in the power grid.

The mathematical model governing the active power-phase angle control of the VSG is denoted as
\begin{equation} \label{eqn7}
    \begin{cases}
        \dfrac{{{\rm{d}}\delta }}{{{\rm{d}}t}} = \omega  - {\omega _0}\\[2mm]
        \dfrac{{{\rm{d}}\omega }}{{{\rm{d}}t}} = \dfrac{1}{J}\left[\dfrac{{{P_{{\rm{ref}}}} - P}}{{{\omega _0}}} - \left({K_D} + \dfrac{{{K_\omega }}}{{{\omega _0}}}\right)\left(\omega  - {\omega _0}\right)\right]
    \end{cases}
\end{equation}
where $\delta$ denotes the virtual power angle (i.e., the phase angle difference between the inverter's GCP voltage and the grid voltage vector), $K_D$ represents the virtual damping coefficient, $J$ indicates the virtual moment of inertia, and $K_\omega$ is the additional primary frequency regulation coefficient.

The reactive power and voltage dynamics governed by the VSG control are mathematically formulated as
\begin{equation} \label{eqn8}
    \begin{cases}
        \dfrac{{{\rm{d}}E}}{{{\rm{d}}t}} = {K_u}({v_{d{\rm{ref}}}} - {v_d}) + {K_q}({Q_{{\rm{ref}}}} - Q)\\
        {E_{{\rm{ref}}}} \!=\! {v_{d{\rm{ref}}}} \!+\! {K_{pQ}}\left[{K_u}({v_{d{\rm{ref}}}} \!-\! {v_d}) \!+\! {K_q}({Q_{{\rm{ref}}}} \!-\! Q)\right] \!+\! {K_{iQ}}E
    \end{cases}
\end{equation}
where $E$ is the integral state variable of the reactive power controller, $K_u$ represents the additional primary voltage regulation coefficient, $K_q$ denotes the reactive power weight factor, $v_{d\mathrm{ref}}$ indicates the $d$-axis reference of the inverter's GCP voltage, $E_{\mathrm{ref}}$ signifies the voltage magnitude reference, and ${K_{pQ}}$ and ${K_{iQ}}$ correspond to the proportional and integral gains of the VSG reactive power-voltage loop (RVL), respectively.

The OVL of the GFM inverter processes the voltage deviations through PI controllers to precisely track the voltage magnitude reference generated by the VSG control block. This loop outputs the $d$- and $q$-axis components of the inverter's AC-side current references. Subsequently, the ICL tracks these current references to generate the corresponding AC-side voltage references in the $dq$ frame. To mitigate the inherent cross-coupling effects induced by the filter, feedforward decoupling terms are incorporated into both the OVL and the ICL. The mathematical model of this voltage-current dual-loop control scheme is expressed as
\begin{equation} \label{eqn9}
    \begin{cases}
        \dfrac{{{\rm{d}}{\gamma _d}}}{{{\rm{d}}t}} = {E_{{\rm{ref}}}} - {v_d}, \quad \dfrac{{{\rm{d}}{\gamma _q}}}{{{\rm{d}}t}} = {v_{q{\rm{ref}}}} - {v_q}\\[2mm]
        {i_{Ld{\rm{ref}}}} = {i_d} - \omega {C_f}{v_q} + {K_{po2}}\left( {{E_{{\rm{ref}}}} - {v_d}} \right) + {K_{io2}}{\gamma _d}\\[2mm]
        {i_{Lq{\rm{ref}}}} = {i_q} + \omega {C_f}{v_d} + {K_{po2}}\left( {{v_{q{\rm{ref}}}} - {v_q}} \right) + {K_{io2}}{\gamma _q}\\[2mm]
        \dfrac{{{\rm{d}}{\xi _d}}}{{{\rm{d}}t}} = {i_{Ld{\rm{ref}}}} - {i_{Ld}}, \quad \dfrac{{{\rm{d}}{\xi _q}}}{{{\rm{d}}t}} = {i_{Lq{\rm{ref}}}} - {i_{Lq}}\\[2mm]
        {e_{d{\rm{ref}}}} = {v_d} - \omega {L_f}{i_{Lq}} + {K_{pi2}}\left( {{i_{Ld{\rm{ref}}}} - {i_{Ld}}} \right) + {K_{ii2}}{\xi _d}\\[2mm]
        {e_{q{\rm{ref}}}} = {v_q} + \omega {L_f}{i_{Ld}} + {K_{pi2}}\left( {{i_{Lq{\rm{ref}}}} - {i_{Lq}}} \right) + {K_{ii2}}{\xi _q}
    \end{cases}\!\!\!\!\!\!\!\!\!\!
\end{equation}
where $v_{q\mathrm{ref}}$ denotes the $q$-axis reference of the inverter's GCP voltage; ${K_{po2}}$ and ${K_{io2}}$ represent the proportional and integral gains of the OVL, respectively; ${K_{pi2}}$ and ${K_{ii2}}$ indicates the proportional and integral gains of the ICL, respectively.

\subsection{Small-signal State-space Switched System Model}
Combining the mathematical models derived in \eqref{eqn1}-\eqref{eqn9} yields the complete nonlinear model of the GCIs under GFL and GFM switching control. By applying small-signal linearization around a steady-state operating point, the full-order small-signal state-space model is obtained as
\begin{equation} \label{eqn10}
    \Delta \dot{\bm{x}} = \bm{A}\Delta \bm{x} + \bm{B}\Delta \bm{u}
\end{equation}
where $\Delta$ denotes the small-signal perturbation; $\bm{x} \in \mathbb{R}^n$ represents the system state vector with respect to time; $\bm{u} \in \mathbb{R}^m$ signifies the control input vector; $\bm{A} \in \mathbb{R}^{n\times n}$ and $\bm{B} \in \mathbb{R}^{n\times m}$ correspond to the system state matrix and input matrix, respectively; $n$ and $m$ indicate the number of state variables and input variables, respectively.

As introduced in literature \cite{Liberzon_switchingsystemscontrol_2003,Lai_dynamicmodelingstability_2025}, a typical switched system contains a collection of continuous-time subsystems and associated switching laws that orchestrates the transitions among these subsystems. The dynamic evolution process of the switched system is determined by the continuous dynamics of active subsystems and the discrete switching events.

Specifically, the GFL subsystem is formulated as a 12th-order state-space model with ${\bm{x}_1} = \left[ \zeta,\delta,{\gamma _d},{\gamma _q},{\xi _d},{\xi _q},{i_d},{i_q},\right.$ \\ $\left. {i_{Ld}},{i_{Lq}},{v_d},{v_q} \right]^{\rm{T}}$ and ${\bm{u}_1} = {\left[ {{P_{{\rm{ref}}}},{Q_{{\rm{ref}}}}} \right]^{\rm{T}}}$. The GFM subsystem is characterized by a 13th-order state-space model with ${\bm{x}_2} = {[\delta ,\omega ,E,{\gamma _d},{\gamma _q},{\xi _d},{\xi _q},{i_d},{i_q},{i_{Ld}},{i_{Lq}},{v_d},{v_q}]^{\rm{T}}}$ and ${\bm{u}_2} = {\left[ {{P_{{\rm{ref}}}},{Q_{{\rm{ref}}}},{v_{d{\rm{ref}}}},{v_{q{\rm{ref}}}}} \right]^{\rm{T}}}$. The design and analysis of the specific switching laws between the GFL and GFM subsystems will be thoroughly discussed in Section IV.

In summary, the small-signal state-space switched system model of GCIs under GFL and GFM switching control is obtained as
\begin{equation} \label{eqn11}
    \!\begin{cases}
		\dot{\bm{x}}_\sigma = {\bm{A}_\sigma}\bm{x}_\sigma+{\bm{B}_\sigma}{\bm{u}_\sigma}, \ \sigma \in \{1,2\} \\[1mm]
		\{\bm{A}_1, \bm{x}_1, \bm{B}_1, \bm{u}_1\} \! \xrightleftharpoons{\text{Switching Laws}} \! \{\bm{A}_2, \bm{x}_2, \bm{B}_2, \bm{u}_2\}
    \end{cases}
\end{equation}
where $\sigma \in \{1, 2\}$ denotes the switching signal indicating the activated subsystem, which is governed by the designed switching strategy. For brevity, the detailed expressions for the system matrices $\bm{A}_1$, $\bm{B}_1$, $\bm{A}_2$ and $\bm{B}_2$ are provided in our open-source document \cite{Lai_Energystorage_2025}.


\section{SSSR Analysis for GFL-GFM Switched System}
In this section, the definition and characterization methodology for the SSSR of the switched system are established. Subsequently, the SSSR and ISMD of the GFL-GFM switched system with respect to the concerned parameters are delineated and analyzed.

\subsection{Definition of the SSSR for Switched System}
In general, a power system can be mathematically modeled as a set of parameterized differential-algebraic equations (DAEs). By performing small-signal linearization around a steady-state equilibrium, the system eigenvalues can be computed to assess small-signal stability. Consequently, the SSSR of a power system is defined as the set of operating points within the parameter space where small-signal stability is strictly preserved, while the boundary of this region is delineated by the operating points corresponding to the system’s marginal stability limits \cite{PanYanFei_hanfengdian_2018}.

According to Lyapunov's indirect method, when all eigenvalues of the state matrix $\bm{A}$ have negative real parts, the system is asymptotically stable about its steady-state operating point $\bm{x}_0$. Consequently, the system is deemed to be small-signal stable. Both the state matrix $\bm{A}$ and the corresponding eigenvalue vector $\bm{\lambda}=\left[ \lambda_1, \lambda_2, \cdots, \lambda_n \right ]$ are functions of the system parameter vector $\bm{p} \in \mathbb{R}^l$. Considering an $l$-dimensional parameter space $\mathcal{S}:=\mathrm{span} \{p_1,p_2,\cdots,p_l\}$ spanned from $\bm{p}$, the SSSR is mathematically defined as
\begin{equation} \label{eqn12}
    \bm{\Omega}:=\left\{ \bm{p} \in \mathcal{S} \mid \max_i\left[ \mathrm{Re}\left( \lambda_i(\bm{p}) \right) \right]<0 \right\}
\end{equation}
where $\mathrm{Re}\left( \lambda_i(\bm{p}) \right)$ denotes the real part of the $i$-th eigenvalue.

Accordingly, the boundary of SSSR is characterized as
\begin{equation} \label{eqn13}
    \partial\bm{\Omega}:=\left\{ \bm{p} \in \mathcal{S} \mid -\varepsilon \leq \max_i\left[ \mathrm{Re}\left( \lambda_i(\bm{p}) \right) \right] \leq 0 \right\}
\end{equation}
where $\varepsilon>0$ represents a predefined boundary tolerance, indicating that the rightmost eigenvalue is located critically close to the imaginary axis.

Furthermore, switching events inherently alter the state-space formulation of the system, consequently reshaping its SSSR. The overall SSSR of a switched system is the union of its constituent subsystems' SSSRs. If the SSSR of one subsystem is entirely contained within that of another, the switching control yields no stability enhancement. In contrast, when neither subsystem's SSSR is a subset of the other, the switching action successfully expands the overall SSSR. In summary, incorporating appropriate switching mechanisms effectively enlarges the switched system's SSSR, thereby enhancing its small-signal stability.




\subsection{Characterization of the SSSR for Switched System}
Given the SSSR definition, switched systems are generally characterized as high-dimensional nonlinear systems. Consequently, it is difficult to derive the analytical boundary of the SSSR for a switched system in the closed form. In this paper, a hyperplane-approximation approach \cite{Yangsu_jiyuyu_2012} is employed to delineated the SSSR of each subsystem, thereby constructing the overall SSSR boundary of the switched system.

Both the state matrix and the eigenvalues of a subsystem are dependent on multiple parameters. As the system parameters vary, the steady-state equilibrium point and the characteristic roots may migrate. When the operating point evolves along a predetermined search ray, starting from an initial asymptotically stable point within the parameter space of interest, the dominant characteristic root may migrate toward the imaginary axis. Once the rightmost root reaches the imaginary axis, the system operates exactly on the SSSR boundary. If the operating point proceeds further along this ray, the rightmost root will cross into the right-half complex plane, rendering the system small-signal unstable.

The detailed procedure for iteratively fitting a subsystem's SSSR boundary within an $l$-dimensional parameter space is outlined as follows.

\begin{table}[!htbp]
\setstretch{0.95}
\centering
\renewcommand{\arraystretch}{1.2}
\small
\begin{tabular}{@{}p{1\linewidth}@{}}  
\toprule \toprule
\textit{Procedure for fitting a subsystem's SSSR boundary.} \\
\midrule
\textbf{\textit{Step 1.}} Select an initial stable operating point and define $l$ mutually orthogonal vectors (MOVs) starting from this point.\\
\textbf{\textit{Step 2.}} Search along the forward and backward directions of each MOV to identify the corresponding $2l$ initial boundary points (BPs) according to \eqref{eqn13}.\\
\textbf{\textit{Step 3.}} Connect each group of $l$ adjacent BPs obtained from the $2^l$ MOV combinations to construct an initial linear approximation of the SSSR boundary, which is an $\left(l\!-\!1\right)$-dimensional hyperplane set.\\
\textbf{\textit{Step 4.}} Compute the global $l$-dimensional volume ($V$) enclosed by the currently fitted boundary.\\
\textbf{\textit{Step 5.}} Starting from the centroids of each hyperplane, refine the boundary by locating more accurate BPs along the normal directions of the respective hyperplanes.\\
\textbf{\textit{Step 6.}} For each newly identified BP, calculate the local $l$-dimensional volume ($V_i$) enclosed by this point and its $l$ neighboring BPs on the previous hyperplane. If the volume ratio $V_i/V> \varepsilon_r$, retain the new BP and return to \textbf{\textit{Step 4}} for further refinement. If there are no new BPs satisfying this criterion, proceed to \textbf{\textit{Step 7}}.\\
\textbf{\textit{Step 7.}} Output the converged set of fitted BPs and the closed-form analytical expressions of the approximated boundary (hyperplane set).\\
\bottomrule \bottomrule
\end{tabular}
\end{table}

The proposed procedure for SSSR boundary fitting offers two primary advantages: (1) It avoids the computational complexity of the analytical derivations, rendering it applicable to large-scale power systems; (2) It facilitates efficient online assessment of system security by providing an intuitive geometric representation of the SSSR.

Ultimately, the overall SSSR of the switched system is constructed by integrating the SSSR boundaries of all constituent subsystems and the corresponding switching laws.



\subsection{SSSRs and ISMDs of the GFL-GFM Switched System}
To facilitate clear geometric visualization, this subsection reveals the impacts of control and operational parameters on system stability within two-dimensional parameter spaces. The small-signal stability margin $\mu$ is defined as the orthogonal distance from the system's rightmost eigenvalue $\lambda_{\rm{m}}$ to the imaginary axis, which equals the absolute value of its real part. Prior to designing the switching laws between the GFL and GFM control modes, it is essential to describe the SSSR and the corresponding ISMD for each subsystem. Based on the parameter settings listed in the Appendix, the SSSR boundary is computed according to the procedure presented in Section III-B (applying convergence tolerances of $\varepsilon=$ 0.01 and $\varepsilon_r=$ 0.001). Subsequently, the Monte Carlo method is employed to conduct uniform random sampling within the established boundary, explicitly mapping the ISMD across the feasible parameter region. 

\subsubsection{GFL Subsystem}
\begin{figure}[!t]
	\centering
    \vspace{-3mm}
	\subfloat[$SCR=$ 2.0]{
		\includegraphics[width=0.4\columnwidth]{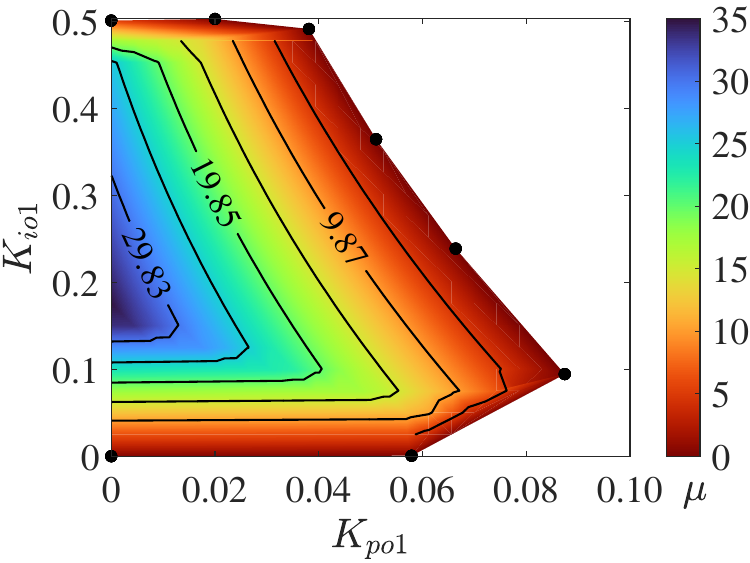}
		\label{Fig2_OPL_SCR2}
	}
	\subfloat[$SCR=$ 4.0]{
		\includegraphics[width=0.4\columnwidth]{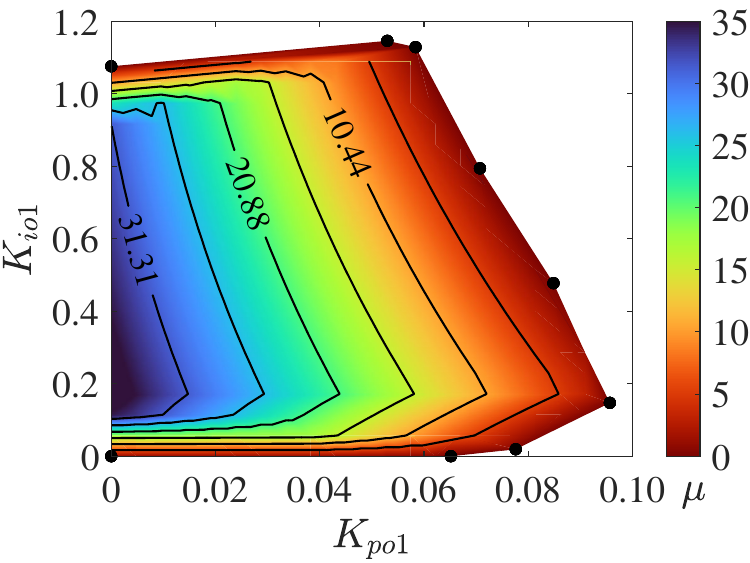}
        \label{Fig2_OPL_SCR4}
	} \\[-2mm]
	\subfloat[$SCR=$ 2.0]{
		\includegraphics[width=0.4\columnwidth]{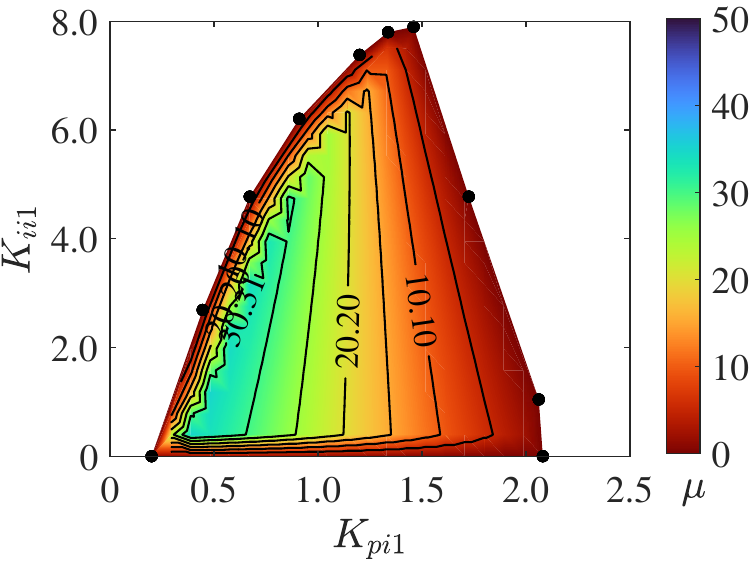}
		\label{Fig2_ICL_SCR2}
	}
	\subfloat[$SCR=$ 4.0]{
		\includegraphics[width=0.4\columnwidth]{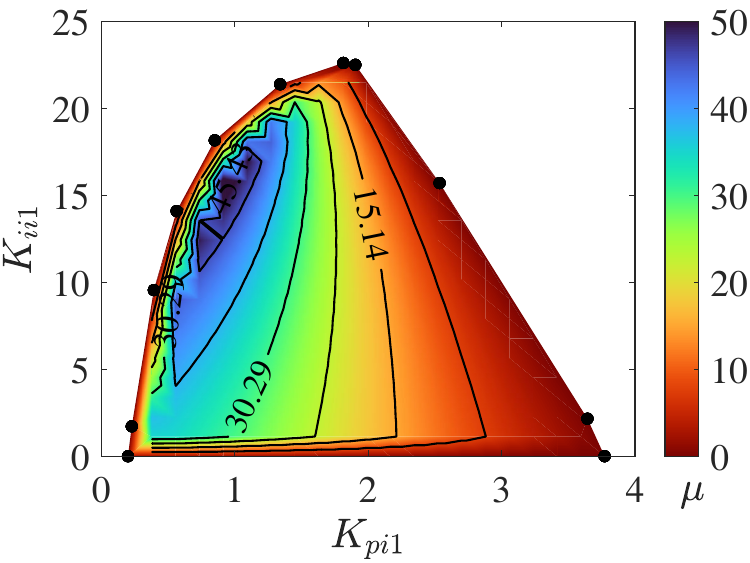}
        \label{Fig2_ICL_SCR4}
	}
	\caption{SSSRs and ISMDs for PI control parameters of the GFL subsystem under different SCRs. (a) and (b) OPL. (c) and (d) ICL. The black dots denote the boundary points. The colormap for $\mu$ indicates the system stability margins, while the solid lines inside SSSR represent stability margin contours.}
    \vspace{-2mm}
	\label{Fig2_PI_GFL}
\end{figure}

\begin{figure}[!t]
	\centering
    \vspace{-3mm}
	\subfloat[$K_{pi1}=$ 1.0]{
		\includegraphics[width=0.4\columnwidth]{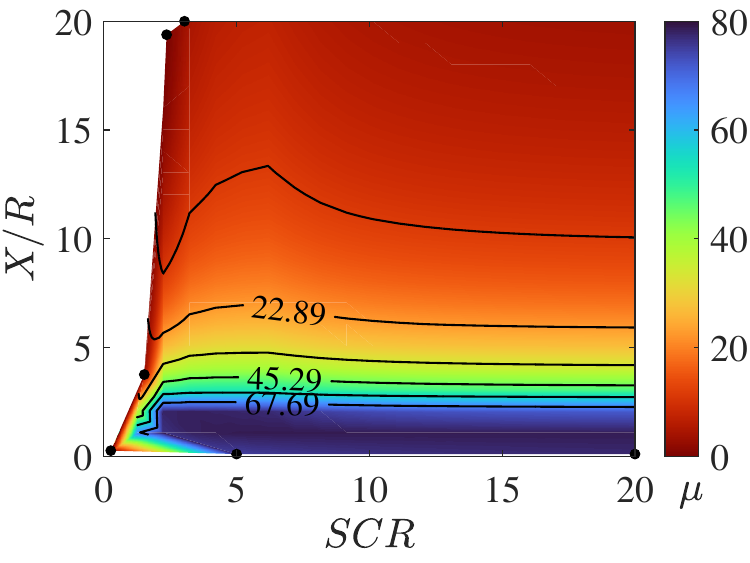}
		\label{Fig3_SCRXR_PI1}
	}
	\subfloat[$K_{pi1}=$ 3.0]{
		\includegraphics[width=0.4\columnwidth]{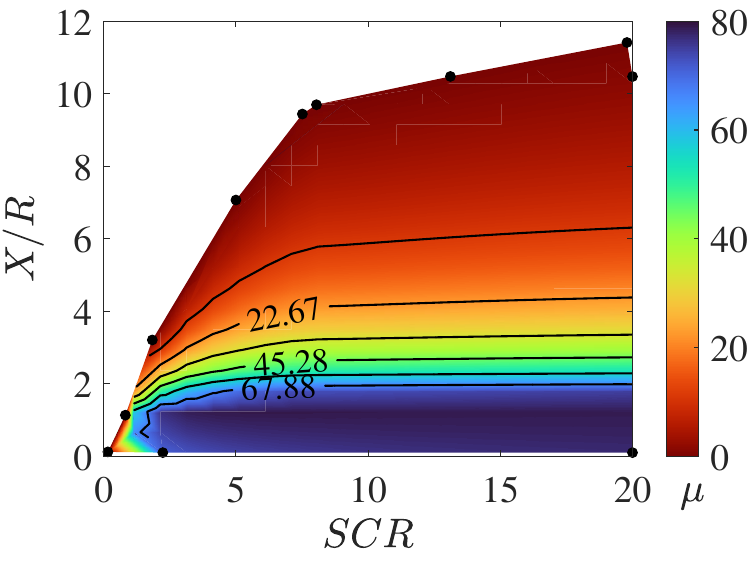}
        \label{Fig3_SCRXR_PI2}
	}
	\caption{SSSRs and ISMDs for operational parameters of the GFL subsystem under different control parameter settings.}
    \vspace{-2mm}
	\label{Fig3_Operation_GFL}
\end{figure}

Regarding control parameters, the SSSRs associated with the PI parameters of the OPL and ICL under different grid strengths (quantified by the SCRs) are depicted in Fig. \ref{Fig2_PI_GFL}. It can be observed that the geometric shapes of the SSSRs differ significantly across various control loops. For each set of PI parameters, the SSSR shape and the corresponding ISMD are similar under different SCRs. The areas of the SSSRs and the maximum stability margins decrease as the grid weakens (i.e., as the SCR decreases), indicating that the GFL subsystem is more stable under a larger SCR. Notably, the SSSR for PI parameters of the ICL is larger than that of the OPL under identical SCR conditions.

For operational parameters, the SSSRs mapped in the $SCR$-$X/R$ plane under different control parameter settings are illustrated in Fig. \ref{Fig3_Operation_GFL}. It is evident that $X/R$ has a more significant influence on system stability and the GFL subsystem exhibits greater stability at smaller $X/R$ values. When $K_{pi1}$ increases to 3.0, the SSSR contracts while the ISMD remains similar.

\subsubsection{GFM Subsystem}
\begin{figure}[!t]
	\centering
    \vspace{-3mm}
	\subfloat[$SCR=$ 2.0]{
		\includegraphics[width=0.4\columnwidth]{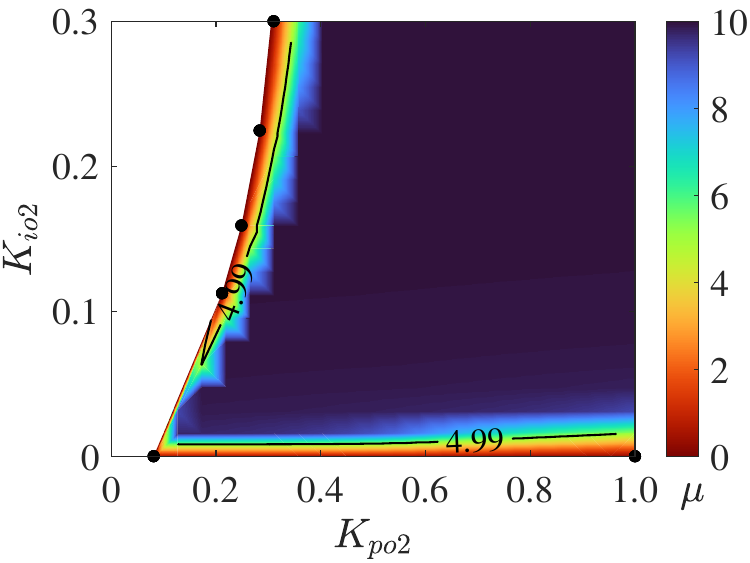}
		\label{Fig4_OVL_SCR2}
	}
	\subfloat[$SCR=$ 4.0]{
		\includegraphics[width=0.4\columnwidth]{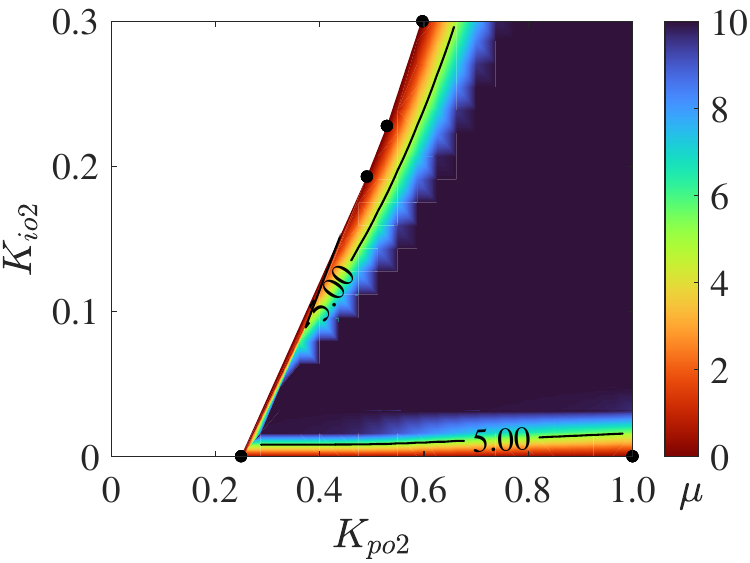}
        \label{Fig4_OVL_SCR4}
	} \\[-2mm]
	\subfloat[$SCR=$ 2.0]{
		\includegraphics[width=0.4\columnwidth]{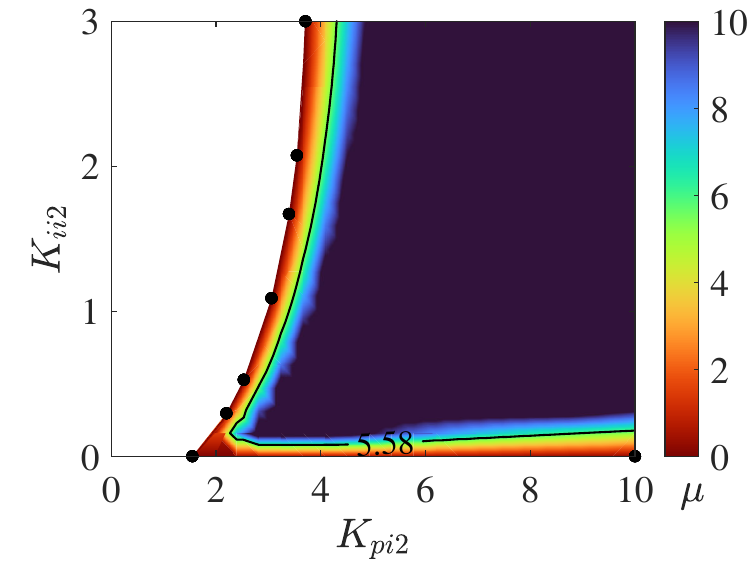}
		\label{Fig4_ICL_SCR2}
	}
	\subfloat[$SCR=$ 4.0]{
		\includegraphics[width=0.4\columnwidth]{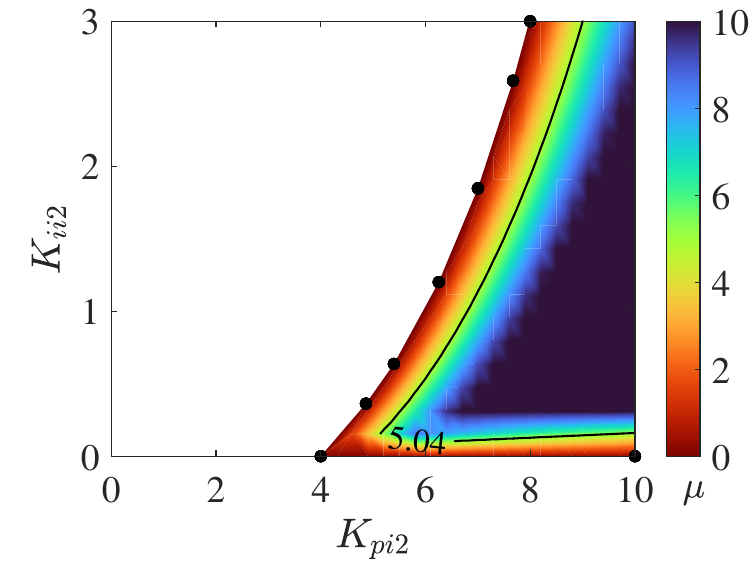}
        \label{Fig4_ICL_SCR4}
	}
	\caption{SSSRs and ISMDs for PI control parameters of the GFM subsystem under different SCRs. (a) and (b) OVL. (c) and (d) ICL.}
    \vspace{-2mm}
	\label{Fig4_PI_GFM}
\end{figure}

\begin{figure}[!t]
	\centering
    \vspace{-3mm}
	\subfloat[$K_{pi2}=$ 10.0]{
		\includegraphics[width=0.4\columnwidth]{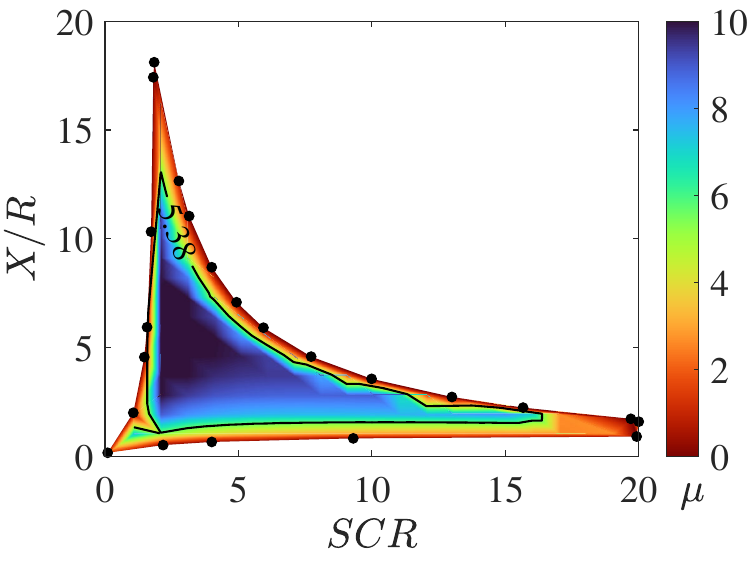}
		\label{Fig5_SCRXR_PI1}
	}
	\subfloat[$K_{pi2}=$ 3.0]{
		\includegraphics[width=0.4\columnwidth]{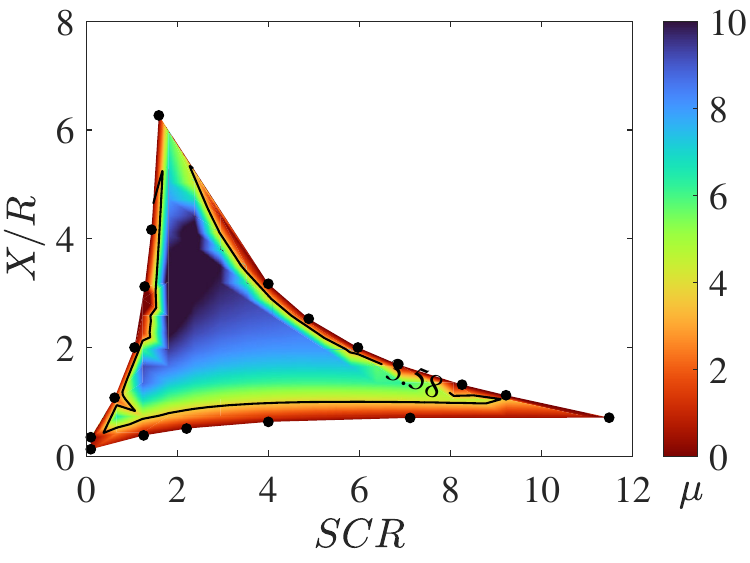}
        \label{Fig5_SCRXR_PI2}
	}
	\caption{SSSRs and ISMDs for operational parameters of the GFM subsystem under different control parameter settings.}
    \vspace{-2mm}
	\label{Fig5_Operation_GFM}
\end{figure}

Regarding control parameters, the SSSRs associated with the PI parameters of the OVL and ICL under different grid strengths are depicted in Fig. \ref{Fig4_PI_GFM}. It can be seen that the geometric shapes of the SSSRs and ISMDs for PI parameters of the voltage-current dual-loop are similar across different SCRs. The areas of the SSSRs decrease as the grid becomes stronger (i.e., as the SCR increases), while the maximum stability margins remain largely unchanged, indicating that the GFM subsystem exhibits greater security under a smaller SCR. Under identical SCR conditions, the SSSR for the PI parameters of the OVL is larger than that of the ICL. For the OVL, the system stability margin will decrease very slowly when $K_{po2}$ or $K_{io2}$ continues to grow. For the ICL, the system becomes more stable as $K_{pi2}$ increases, while it slowly drifts toward instability with the growth of $K_{ii2}$.

For operational parameters, the SSSRs mapped in the $SCR$-$X/R$ plane under different control parameter settings are illustrated in Fig. \ref{Fig5_Operation_GFM}. Both the SCR and $X/R$ need to be maintained at low values to ensure system stability. When $K_{pi2}$ is decreased to 3.0, both the SSSR area and the stability margin become smaller.

\section{Switching Control Strategy Design for the GFL-GFM Switched System}
Based on the established SSSR and ISMD characterization methodology, this section utilizes a GMM to derive the closed-form analytical expressions for the system stability margin and parameter sensitivity. Subsequently, a novel CSI integrating the stability margin, parameter sensitivity, and boundary distance indicators is proposed. Furthermore, considering multiple system parameters and performance indicators, a CSI-driven adaptive switching control strategy for the GFL-GFM switched system is developed.

\subsection{GMM-based Stability Margin and Sensitivity Estimation}
The SSSRs of both subsystems are generated using a series of multi-dimensional scatter points obtained from Monte Carlo sampling. Given the inherent high-dimensional correlations between the system parameters and stability margins, their relationship can be accurately captured by a joint probability distribution. Therefore, a GMM is employed to fit these scatter points, yielding closed-form analytical expressions for the parameter-dependent stability margin and its associated sensitivity. Based on this stability margin and sensitivity information, the optimal operating conditions can be determined analytically.

The GMM is widely used to describe the joint probability distribution of a $d$-dimensional random vector $\bm{V} \in \mathbb{R}^d$. It is defined as a convex combination of $K$ Gaussian components with corresponding weights $\omega_k$, formulated as
\begin{subequations} \label{Eqn14}
		\begin{gather}
			p(\bm{V})=\sum_{k=1}^K \omega_k \mathcal{N}_k\left(\bm{V} \mid \bm{\mu}_k, \bm{\Sigma}_k\right) \label{Eqn14a} \\[1mm]
			\sum_{k=1}^K \omega_k=1, \quad \omega_k>0 \label{Eqn14b} \\[1mm]
			\mathcal{N}_k\left(\bm{V} \mid \bm{\mu}_k, \bm{\Sigma}_k\right) = \dfrac{\mathrm{e}^{-\frac{1}{2}\left(\bm{V}-\bm{\mu}_k\right)^{\T} \bm{\Sigma}_k^{-1}\left(\bm{V}-\bm{\mu}_k\right)}} {(2\pi)^{d/2} |\bm{\Sigma}_k|^{1/2}} \label{Eqn14c}
		\end{gather}
	\end{subequations}
where $p\left( \bm{V} \right)$ represents the joint probability density function of $\bm{V}$; $\mathcal{N}_k$ denotes the multivariate normal distribution of the $k$-th Gaussian component; $\omega_k$, $\bm{\mu}_k$, and $\bm{\Sigma}_k$ are its corresponding weight, mean vector, and covariance matrix, respectively.

Consider a set of $(d+1)$-dimensional data points $\left(x_1,\cdots,x_d,y\right)$, where the input vector $\bm{X} = \left[x_1,\cdots,x_d\right]^{\mathrm{T}} \in \mathbb{R}^d$ represent the parameters of subsystems, and the scalar output $Y = y \in \mathbb{R}$ denotes the corresponding stability margin under different parameter settings. By employing the GMM defined in \eqref{Eqn14}, the joint probability density function of these variables can be expressed as
\begin{equation}
    p\left(\begin{bmatrix}
            \bm{X} \\ Y
        \end{bmatrix}\right) = \sum_{k=1}^{K} \omega_k \, \mathcal{N}_k
    \left(\begin{bmatrix}
            \bm{X} \\ Y
        \end{bmatrix} \middle|
        \begin{bmatrix}
            \bm{\mu}_{\bm{X}} \\ \mu_{Y}
        \end{bmatrix}_k,
        \begin{bmatrix}
            \bm{\Sigma}_{\bm{X}\bm{X}} & \bm{\Sigma}_{\bm{X}Y} \\
            \bm{\Sigma}_{Y\bm{X}} & \Sigma_{YY}
        \end{bmatrix}_k\right).
    \label{Eqn15}
\end{equation}

By applying the expectation maximization algorithm to the dataset, the parameters of the GMM \eqref{Eqn15} are iteratively optimized.  Subsequently, the marginal probability density function of the input vector $\bm{X}$ is derived as
\begin{equation}
    p(\bm{X})= \sum_{k=1}^{K} \omega_k \, \mathcal{N}_k \left( \bm{X} \mid \bm{\mu}_{\bm{X},k}, \bm{\Sigma}_{\bm{X}\bm{X},k} \right).
    \label{Eqn16}
\end{equation}

Based on \eqref{Eqn16}, the posterior responsibility of the $k$-th component given the input vector $\bm{X}$ is derived as
\begin{equation}
    \!\!\!\!\gamma_k(\bm{X}) = p(k\mid\bm{X}) = \dfrac{\omega_k\,\mathcal{N}_k \left( \bm{X}\mid \bm{\mu}_{\bm{X},k}, \bm{\Sigma}_{\bm{X}\bm{X},k} \right)}  {\sum_{j=1}^{K} \omega_j\,\mathcal{N}_j \left( \bm{X}\mid \bm{\mu}_{\bm{X},j}, \bm{\Sigma}_{\bm{X}\bm{X},j} \right)}.\!\!\!\!
    \label{Eqn17}
\end{equation}

Moreover, the conditional expectation of $Y$
given $\bm{X}$ within the $k$-th component is expressed as
\begin{equation}
    m_k(\bm{X}) = \mathbb{E}[Y\mid\bm{X},k] = \mu_{Y,k} + \bm{\Sigma}_{Y\bm{X},k} \bm{\Sigma}_{\bm{X}\bm{X},k}^{-1} \left(\bm{X}-\bm{\mu}_{\bm{X},k}\right).
    \label{Eqn18}
\end{equation}

Combining \eqref{Eqn17} and \eqref{Eqn18}, the total conditional expectation of the stability margin $Y$ is obtained as
\begin{equation}
    Y = f(\bm{X}) = \mathbb{E}\left[Y\mid\bm{X}\right] = \sum_{k=1}^{K}  \gamma_k(\bm{X}) m_k(\bm{X}).
    \label{Eqn19}
\end{equation}

Formula \eqref{Eqn19} establishes a smooth $(d+1)$-dimensional regression hypersurface over the sampled data points, effectively approximating the high-dimensional nonlinear mapping from the system parameters $\left(x_1,\cdots,x_d\right)$ to the stability margin $y$.

Furthermore, the parameter sensitivity can be quantified by the gradient of $f(\bm{X})$, which is derived as
\begin{equation}
    \nabla f(\bm{X}) = \sum_{k=1}^{K} \left[ \nabla\gamma_k(\bm{X})\,m_k(\bm{X})
    + \gamma_k(\bm{X})\, \nabla m_k(\bm{X}) \right]
    \label{Eqn20}
\end{equation}
where the gradient of the posterior responsibility $\nabla\gamma_k$ and the gradient of the conditional expectation $\nabla m_k$ are respectively given by
\begin{equation}
    \begin{aligned}
        &\nabla\gamma_k =
        \dfrac{\left( \sum_{j=1}^{K} \omega_j\,\mathcal{N}_j \right)\nabla \left(\omega_k\,\mathcal{N}_k\right) - \left(\omega_k\,\mathcal{N}_k\right) \nabla\left(\sum_{j=1}^{K} \omega_j\,\mathcal{N}_j \right)}{\left( \sum_{j=1}^{K} \omega_j\,\mathcal{N}_j \right) ^{2}} \\
        & =
        \gamma_k
        \left[
        - \bm{\Sigma}_{\bm{X}\bm{X},k}^{-1} \left(\bm{X} - \bm{\mu}_{\bm{X},k} \right)
        + \sum_{j=1}^{K} \gamma_j \bm{\Sigma}_{\bm{X}\bm{X},j}^{-1} \left(\bm{X} - \bm{\mu}_{\bm{X},j} \right)
        \right],
    \end{aligned}
    \label{Eqn21}
\end{equation}
and
\begin{equation}
\nabla m_k
= \left( \bm{\Sigma}_{Y\bm{X},k}\bm{\Sigma}_{\bm{X}\bm{X},k}^{-1} \right)^\T.
\label{Eqn22}
\end{equation}

The sensitivity vector $\nabla f(\bm{X}) = [\partial f / \partial x_1,\cdots, \partial f / \partial x_d]^{\mathrm{T}}$ quantifies the local change rate of the GMM-derived $(d+1)$-dimensional regression hypersurface with respect to the input coordinates $\left(x_1,\cdots,x_d\right)$. This analytical gradient accurately approximates the sensitivity of the system's stability margin to variations in parameters, providing a rigorous mathematical foundation for evaluating the robustness of operating points.

\subsection{Comprehensive Stability Index}

Due to the highly nonlinear dynamic characteristics in the parameter space of GCI systems, a simple stability margin index is often insufficient to comprehensively reflect the system's dynamic security properties. Specifically, operating points with large stability margins may nevertheless lack robustness if they are located in highly sensitive regions or in close geometric proximity to the SSSR boundaries, rendering them vulnerable to parameter fluctuations. To quantitatively evaluate the overall stability of an operating point $\bm{X}$, this paper proposes a CSI that integrates the stability margin, parameter sensitivity, and distance to the security boundary. This index leverages the stability margin and gradient expressions derived from the GMM formulations in \eqref{Eqn19} and \eqref{Eqn20}.


To eliminate the impacts of varying physical dimensions and scales, each stability indicator is normalized to a range of [0,1] by using the max-min normalization. Synthesizing these normalized indicators, the CSI is defined as a dimensionless comprehensive evaluation function, which is formulated as
\begin{equation}
J(\bm{X}) = \omega_m \cdot \bar{M}(\bm{X}) - \omega_s \cdot \bar{S}(\bm{X}) + \omega_d \cdot \bar{D}(\bm{X}) 
\label{Eqn26}
\end{equation}
where $\bm{X}$ denotes the system parameter vector and $\omega_m, \omega_s, \omega_d$ represent the weighting coefficients for each normalized indicator, satisfying $\omega_m + \omega_s + \omega_d =$ 1 and $\omega_m, \omega_s, \omega_d \in$ [0,1]. By tuning these weight combinations, this formulation facilitates a flexible trade-off between stability and robustness based on practical engineering requirements. In \eqref{Eqn26}, the normalized indicators are introduced as follows:
\begin{itemize}
    \item[$\bullet$] The normalized stability margin $\bar{M}(\bm{X})$, which is derived from the GMM expectation $f(\bm{X})$ in \eqref{Eqn19}, quantifies the damping reserve that maintains the system at a secure distance from the critical instability boundary. Specifically, a larger value of $\bar{M}(\bm{X})$  indicates a faster convergence rate of the system state variables following a disturbance.
    \item[$\bullet$] The normalized parameter sensitivity $\bar{S}(\bm{X})$ quantifies the vulnerability of the system's stability margin to parameter variations. It is evaluated based on the Euclidean norm of the gradient vector $\nabla f(\bm{X})$ obtained in \eqref{Eqn20}. This indicator serves as a penalty term within the CSI, where a smaller sensitivity value signifies that the operating point possesses greater robustness against parameter drift.
    \item[$\bullet$] The normalized boundary distance $\bar{D}(\bm{X})$ represents the minimum Euclidean distance from the operating point to the nearest SSSR boundary within the parameter space. This metric explicitly quantifies the fault tolerance of the system under large parameter disturbances, where a greater distance indicates a larger security margin.
\end{itemize}

Consequently, the proposed CSI effectively identifies the superior operating points that possess high stability margins, exhibit low sensitivity to parameter fluctuations, and reside near the geometric center of the SSSR.

\subsection{Switching Control Strategy for GFL-GFM Switched System}
Existing studies predominantly rely on the SCR as the switching indicator between the GFL and GFM control modes, largely neglecting the coupled impacts of variations in other system parameters. Furthermore, these conventional switching strategies typically trigger a transition when the system is on the verge of a subsystem's security boundary. Consequently, the system becomes highly vulnerable to perturbations prior to the switching event, thereby lacking adequate robustness. To address this critical gap, the proposed CSI is employed to characterize the stability and robustness of an operating point, fully reflecting the influence of multiple parameters.

\begin{figure}[!t]
	\centering
    \vspace{-3mm}
    \includegraphics[width=0.6\columnwidth]{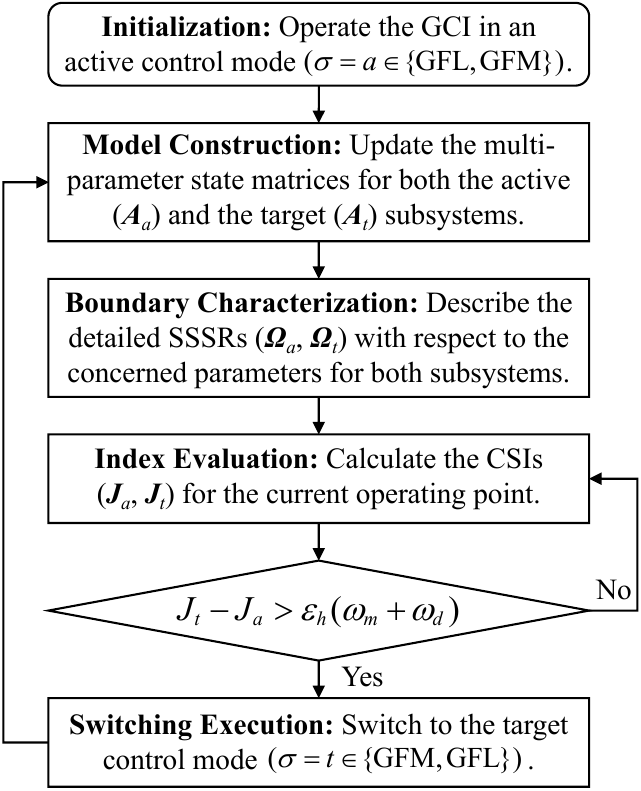}
	\caption{Flowchart of the proposed CSI-based switching control strategy. The parameter $\varepsilon_h$ represents the hysteresis threshold, and the sum term $(\omega_m+\omega_d)$ denotes the theoretical maximum value of the CSI.}
    \vspace{-2mm}
	\label{Fig6_Switching_flow}
\end{figure}

From the perspective of the CSI, three fundamental switching principles are established for the GFL-GFM switched system: (1) The SSSR of one subsystem must not be entirely contained by that of the other; (2) A switching event is triggered when the CSI of the target subsystem exceeds that of the currently active one; (3) A hysteresis switching mechanism should be implemented to prevent switching oscillations. By employing these principles, the overall SSSR of the switched system can be effectively expanded, enhancing operational security under varying grid conditions. The flowchart of the proposed switching control strategy is illustrated in Fig. \ref{Fig6_Switching_flow}.

\section{Case Study}
Based on the GCI architecture and mathematical models formulated in Section II, a structured EMT simulation model for the GFL-GFM switchable inverter connected to an infinite bus is developed on the CloudPSS platform \cite{Song_CloudPSS_2016,Song_cloudpsshighperformancepower_2020,Chen_mianxiangxin_2022}. To ensure broad applicability, this model employs averaged representations of switching devices and is normalized using per-unit values with single-machine equivalence scaling. This modeling approach facilitates seamless adaptation to equipment with diverse rated voltages and power capacities, as well as to device clusters and plant-level equivalents. Furthermore, the complete EMT test model has been made open-source and is accessible in \cite{Lai_GFLGFMmodel_2016}.

In this section, the effectiveness of the proposed GFL-GFM switched system model, the SSSR description and parameter sensitivity analysis approach, and the CSI-based GFL-GFM switching control strategy are demonstrated through case studies integrated Matlab computations and CloudPSS simulations.

\subsection{Verification of GFL-GFM Switched System Model}
This subsection evaluates the fidelity of the developed GFL and GFM inverter models by benchmarking the state-space theoretical calculations performed in Matlab against the time-domain EMT simulations  conducted on CloudPSS. The consistency between the state-space representations and the EMT nodal analysis serves as a rigorous validation of the proposed modeling framework.

\begin{figure}[!t]
	\centering
    \vspace{-2.5mm}
	\subfloat[GFL Subsystem]{
		\includegraphics[width=0.6\columnwidth]{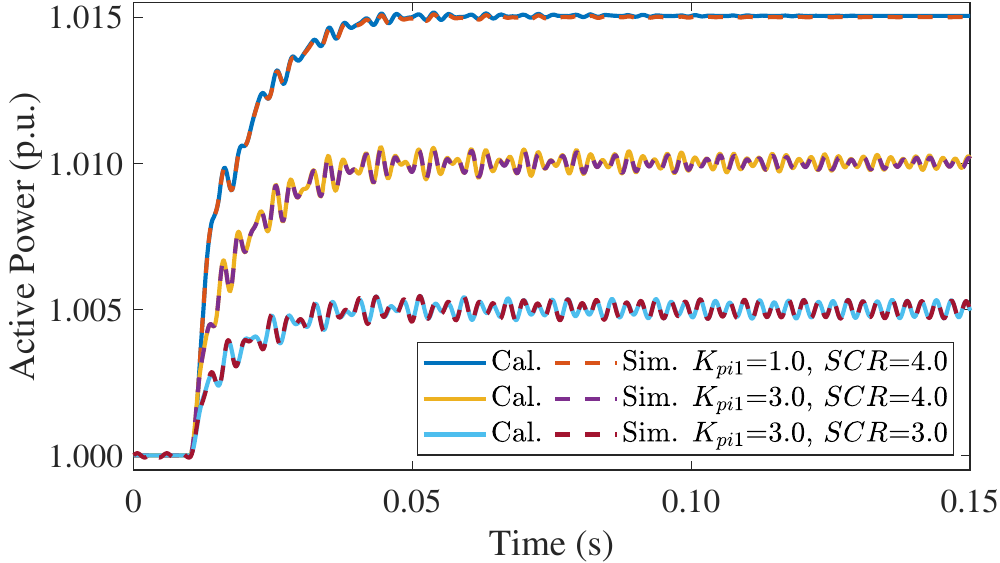}
		\label{Fig7_GFL_Veri1}
	}\\[-0.2mm]
	\subfloat[GFM Subsystem]{
		\includegraphics[width=0.6\columnwidth]{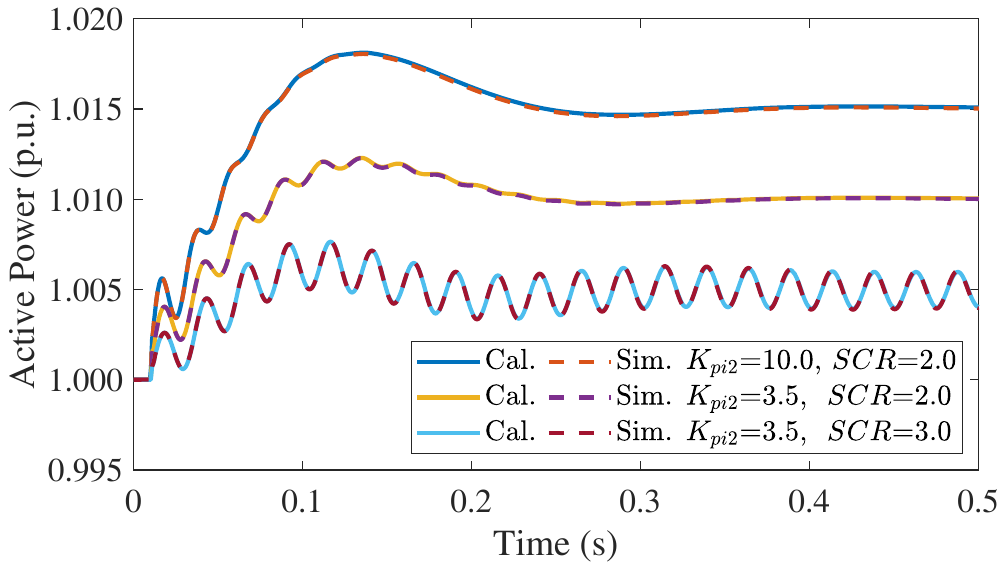}
		\label{Fig7_GFM_Veri1}
	}
	\caption{Verification of the GFL-GFM switched system model: Comparison of the active power responses between state-space theoretical calculations (solid lines) and EMT simulations (dashed lines) under small-signal disturbances.}
    \vspace{-2mm}
	\label{Fig7_Model_verification}
\end{figure}

A single GCI connected to an infinite bus is adopted as the test case, with the detailed parameter configurations provided in the Appendix. Upon reaching the steady state, step disturbances of $\Delta P_{\rm{ref}} =$ 0.005, 0.010, 0.015 p.u. are applied at $t =$ 0.01 s, respectively. Fig. \ref{Fig7_Model_verification} illustrates the comparison of the active power transient response curves between the theoretical calculation and EMT simulation results across various system parameter settings, where $P={v_d}{i_d}+{v_q}{i_q}$.

As depicted in Fig. \ref{Fig7_Model_verification}, for system operating points with different stability margins, the small-signal responses obtained from state-space  analytical models and EMT simulation models for both GFL and GFM subsystems exhibit a high degree of congruence. Specifically, the root-mean-square errors remain below $\text{2}\times\text{10}^{-\text{4}}$ in all cases, confirming the accuracy of the proposed GFL–GFM switched system model. Notably, while both subsystems converge to the same steady-state equilibrium points under identical small-signal disturbances, the GFL subsystem demonstrates faster transient responses compared to the GFM subsystem.

\subsection{Effectiveness of SSSR and Parameter Sensitivity Analysis}
This subsection verifies the effectiveness of the proposed SSSR and ISMD characterization methods for the GFL and GFM subsystems through EMT simulations conducted across various operating points. Furthermore, the GMM-based estimation approach for stability margin and parameter sensitivity is validated by analyzing the trajectories of the system's dominant eigenvalues and the corresponding variations in parameter sensitivity.

\begin{figure}[!t]
	\centering
	\vspace{-3mm}
	\subfloat[GFL Subsystem]{
		\includegraphics[width=0.6\columnwidth]{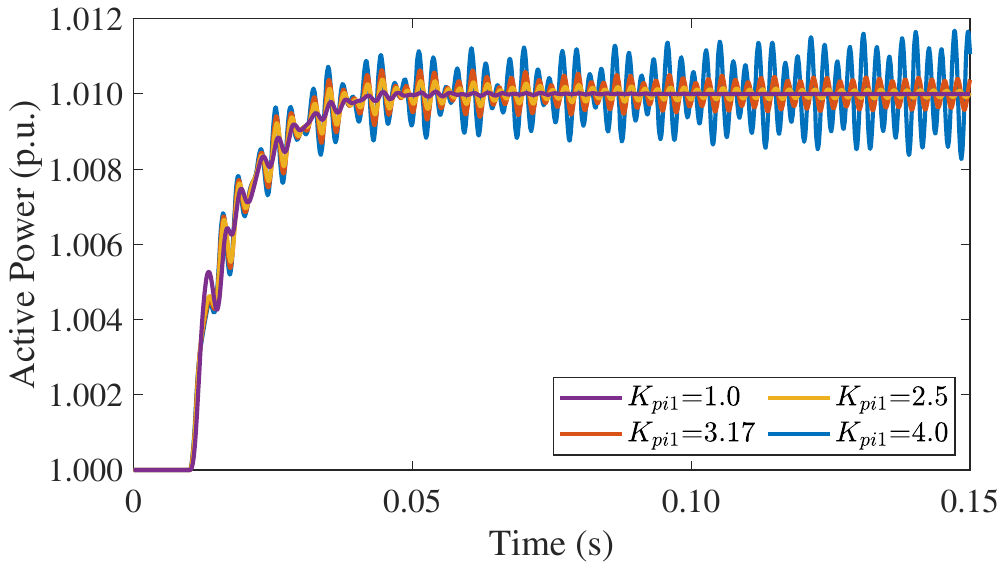}
		\label{Fig8_GFL_Veri2}
	}\\[-0.2mm]
	\subfloat[GFM Subsystem]{
		\includegraphics[width=0.6\columnwidth]{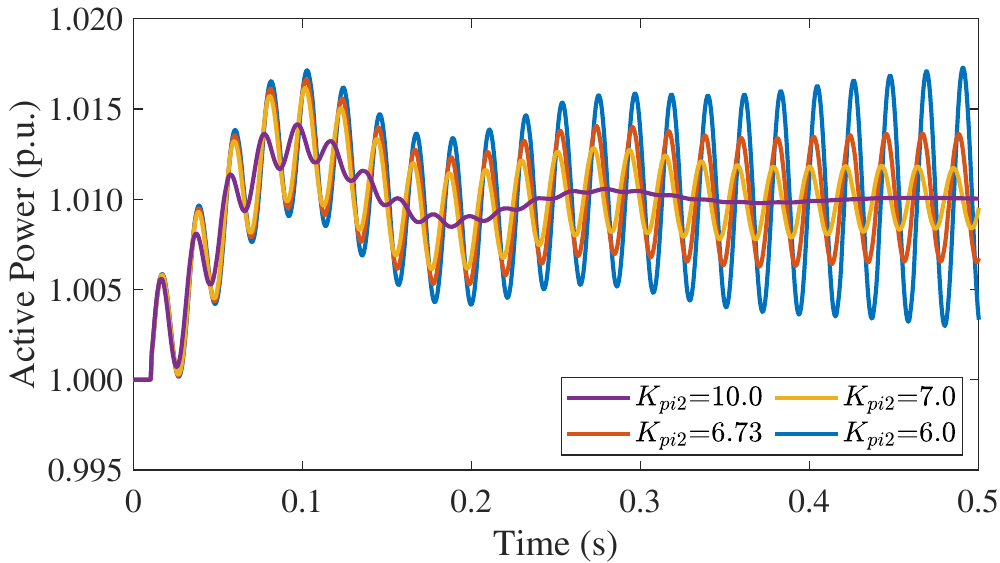}
		\label{Fig8_GFM_Veri2}
	}
	\caption{Verification of the SSSR and ISMD description: Comparison of active power responses at various operating points under different stability conditions following small-signal disturbances.}
    \vspace{-2mm}
	\label{Fig8_SSSR_verification}
\end{figure}

Based on the established SSSRs for the PI parameters of the ICL, four distinct parameter configurations of $(K_{pi},K_{ii})$ are selected for both the GFL and GFM subsystems. These test points, derived from the SSSRs illustrated in Fig. \ref{Fig2_PI_GFL}\subref{Fig2_ICL_SCR2} and Fig. \ref{Fig4_PI_GFM}\subref{Fig4_ICL_SCR4}, respectively, represent four specific stability conditions: (i) deep within the security region, (ii) adjacent to the boundary, (iii) precisely on the boundary, and (iv) outside the security region. The corresponding parameter coordinates selected for the GFL subsystem are (1.0,2500), (2.5,2500), (3.17,2500), and (4.0,2500), while those for the GFM subsystem are (10.0,500), (7.0,500), (6.73,500), and (6.0,500). For these operating points, the EMT simulation results of the active power responses under small-signal disturbances are presented in Fig. \ref{Fig8_SSSR_verification}.

As depicted in Fig. \ref{Fig8_SSSR_verification}, the operating points located within the security region and adjacent to the boundary both exhibit stable responses. Notably, the configuration deep within the security region demonstrates more rapid convergence to the steady state, indicating a larger stability margin. Moreover, the operating point situated exactly on the SSSR boundary is marginally stable, characterized by sustained active power oscillations. Conversely, the operating point outside the security region is unstable, displaying divergent oscillations. These results validate the accuracy of the SSSRs and ISMDs illustrated in Fig. \ref{Fig2_PI_GFL}--Fig. \ref{Fig5_Operation_GFM}.

\begin{figure}[!t]
	\centering
    \vspace{-6.5mm}
	\subfloat[$SCR=$ 2--18, $X/R=$ 10]{
		\includegraphics[width=0.45\columnwidth]{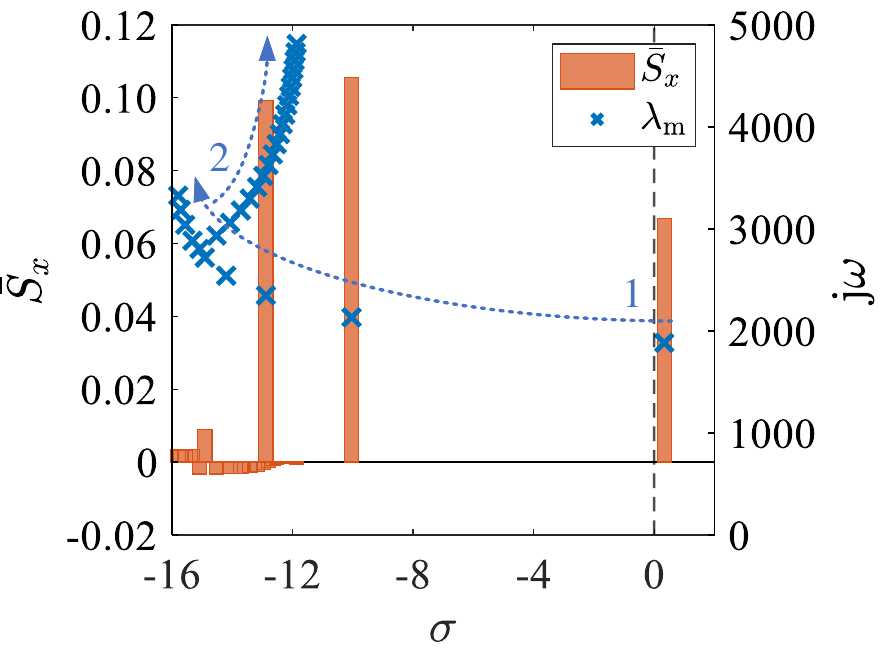}
		\label{Fig9_GFL_Veri3_SCR}
	}
	\subfloat[$SCR=$ 10, $X/R=$ 0--20]{
		\includegraphics[width=0.45\columnwidth]{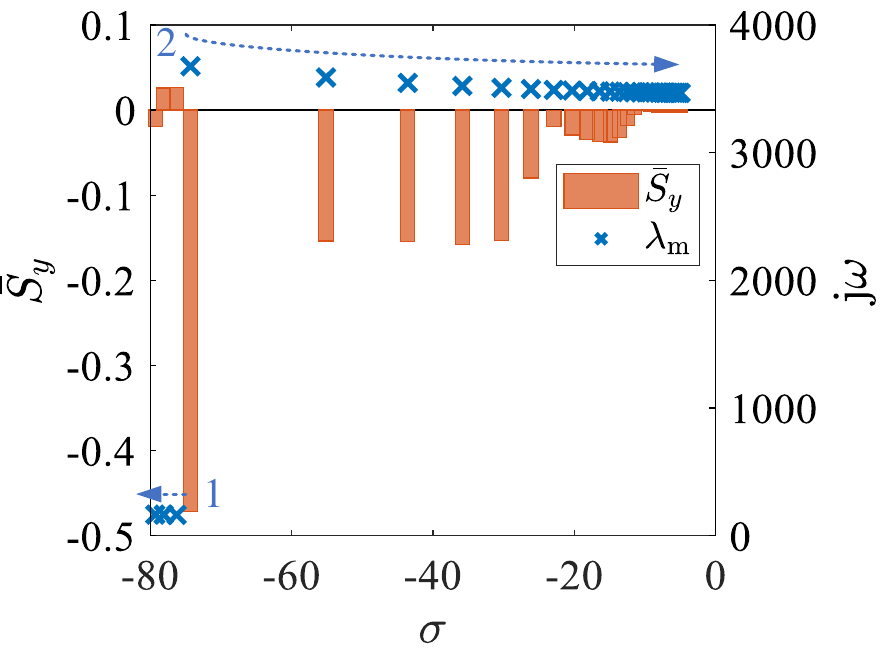}
        \label{Fig9_GFL_Veri3_XR}
	} \\[-2mm]
	\subfloat[$SCR=$ 2--18, $X/R=$ 2]{
		\includegraphics[width=0.45\columnwidth]{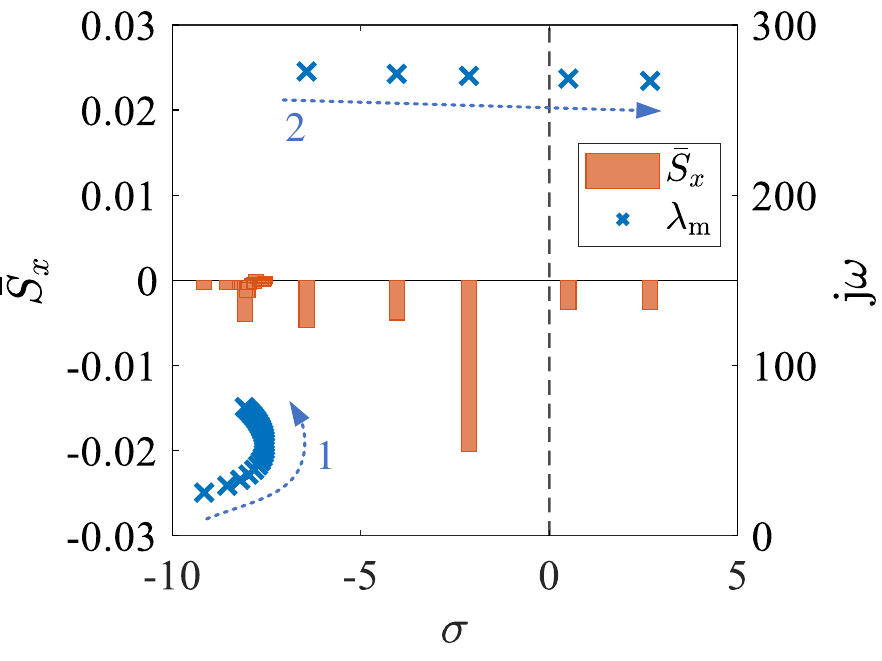}
		\label{Fig9_GFM_Veri3_SCR}
	}
	\subfloat[$SCR=$ 2, $X/R=$ 0--20]{
		\includegraphics[width=0.45\columnwidth]{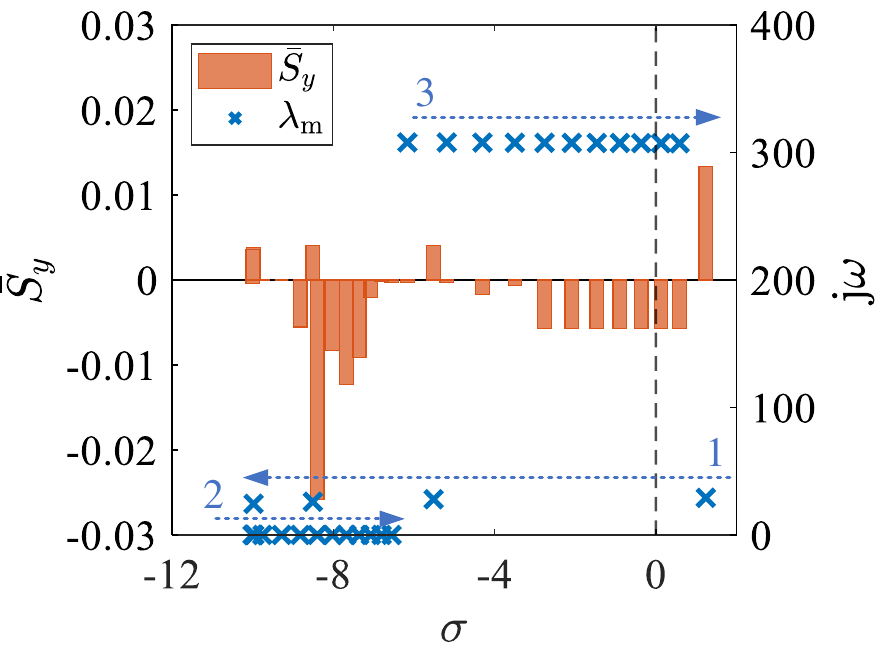}
        \label{Fig9_GFM_Veri3_XR}
	}
	\caption{Validation of the parameter sensitivity estimation: Trajectories of the system’s dominant eigenvalues and the corresponding sensitivity distributions. (a) and (b) GFL subsystem; (c) and (d) GFM subsystem. The arrows indicate the direction of parameter increment, and the numbers denote distinct dominant modes.}
    \vspace{-2mm}
	\label{Fig9_Sensitivity_verification}
\end{figure}

With reference to the SSSRs of the $SCR$ and $X/R$ parameters for the GFL and GFM subsystems as illustrated in Fig. \ref{Fig3_Operation_GFL}\subref{Fig3_SCRXR_PI1} and Fig. \ref{Fig5_Operation_GFM}\subref {Fig5_SCRXR_PI1}, the coefficients of determination ($R^2$) for the ISMD estimation are 0.96 and 0.93, respectively. Furthermore, Fig. \ref{Fig9_Sensitivity_verification} depicts the trajectories of the system’s dominant eigenvalues and the corresponding sensitivity bar charts as $SCR$ or $X/R$ is uniformly increased.

As shown in Fig. \ref{Fig9_Sensitivity_verification}, the system stability margin varies non-monotonically with respect to the $SCR$ and $X/R$ parameters. Specifically, a positive (or negative) sensitivity value corresponds to an increasing (or decreasing) trend in the stability margin, and the spatial density of the eigenvalues along the trajectory correlates inversely with the sensitivity magnitude. For the GFL subsystem, the system stability margin is more sensitive to $X/R$, while the GFM subsystem exhibits comparable sensitivity to both parameters. These results demonstrate the accuracy of the proposed stability margin and parameter sensitivity estimation method.

\subsection{Test of CSI and GFL-GFM Switching Control Strategy}
\begin{figure}[!t]
	\centering
    \vspace{-3.5mm}
	\subfloat[GFL Subsystem]{
		\includegraphics[width=0.4\columnwidth]{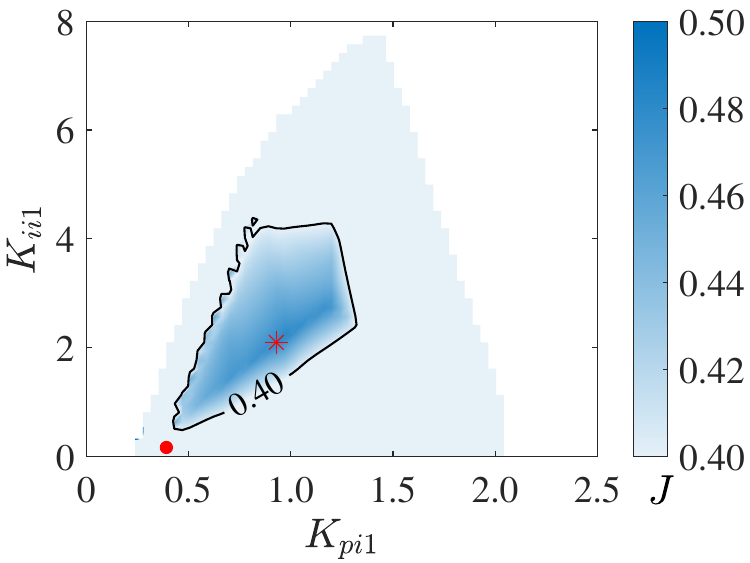}
		\label{Fig10_GFL_Veri4_CIL}
	}
	\subfloat[GFM Subsystem]{
		\includegraphics[width=0.4\columnwidth]{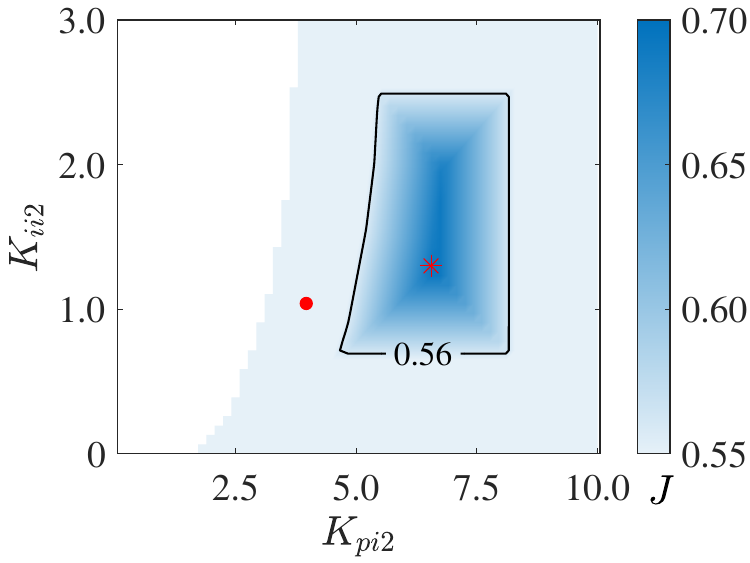}
        \label{Fig10_GFM_Veri4_CIL}
	}
	\caption{CSI distributions for the PI parameters of the ICL under both GFL and GFM control modes. The red star and dot pinpoint the optimal parameter configurations yielding the maximum CSI and the maximum stability margin, respectively. The color gradient reflects the calculated CSI values. The solid contour lines inside the SSSR represent the 80\% threshold of the maximum CSI, indicating a highly robust operating subregion.}
    \vspace{-1.5mm}
	\label{Fig10_CSI_CIL}
\end{figure}

Building upon the SSSRs of the ICL PI parameters for the GFL and GFM subsystems, as illustrated in Fig. \ref{Fig2_PI_GFL}\subref{Fig2_ICL_SCR2} and Fig. \ref{Fig4_PI_GFM}\subref{Fig4_ICL_SCR2}, respectively, the corresponding CSI distributions within these regions are evaluated. By adopting a designated weighting vector of $\bm{\omega}=$ [0.4,0.3,0.3] in \eqref{Eqn26}, the resulting CSI mappings are depicted in Fig. \ref{Fig10_CSI_CIL}.

As revealed in Fig. \ref{Fig10_CSI_CIL}, the GFM subsystem generally exhibits a higher CSI compared to the GFL subsystem when $SCR =$ 2.0. In particular, the parameter configurations that yield the maximum CSI and the maximum stability margin are located at distinctly different operating points. The coordinate point corresponding to the maximum stability margin is situated precariously close to the SSSR boundary, and it is also highly sensitive to parameter variations, thereby lacking adequate robustness. This observation demonstrates that the proposed CSI comprehensively captures the stability and parametric robustness of an operating point, making it superior to a single evaluation metric.

\begin{figure}[!t]
	\centering
	\vspace{-3mm}
	\includegraphics[width=0.7\columnwidth]{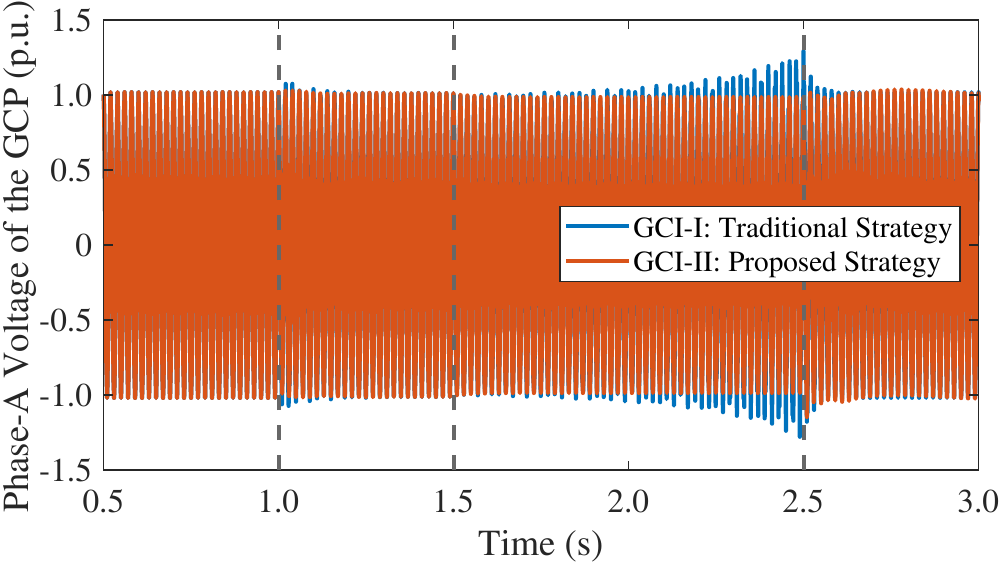}
	\caption{Verification of the CSI-based GFL-GFM switching control strategy: Comparison of the phase-$A$ voltage at the GCP of the GCIs under different switching control strategies and varying operating conditions.}
    \vspace{-2mm}
	\label{Fig11_Veri5_Switching_comp}
\end{figure}

To verify the proposed CSI-driven GFL-GFM switching control strategy, the phase-$A$ voltage waveforms at the GCP of GCIs employing the traditional strategy \cite{Li_gaoshentou_2021} (GCI-I) and the proposed strategy (GCI-II) under varying operating conditions are illustrated in Fig. \ref{Fig11_Veri5_Switching_comp}. The detailed validation process is outlined as follows:
\begin{itemize}
    \item[$\bullet$] \textit{0.5--1.0 s:} Both GCIs operate steadily within the GFL control mode under a strong grid condition ($SCR=$ 6.0).
    \item[$\bullet$] \textit{1.0--1.5 s:} At 1.0 s, the grid strength drops to an SCR of 3.1. Consequently, GCI-II transitions to the GFM control mode, whereas the traditional GCI-I takes no switching action. While both GCIs maintain stability, GCI-I operates perilously close to its SSSR boundary.
    \item[$\bullet$] \textit{1.5--2.5 s:} Following a subsequent parameter disturbance at 1.5 s (e.g., the $X/R$ increases to 8.0), GCI-II remains stable. In contrast, the voltage waveform of GCI-I diverges, demonstrating a loss of small-signal stability.
    \item[$\bullet$] \textit{2.5--3.0 s:} At 2.5 s, as the grid strength recovers to an SCR of 7.0, GCI-II adaptively switches back to the GFL control mode, and GCI-I regains stable operation.
\end{itemize}

In summary, the proposed switching mechanism effectively enhances system stability and robustness under variations in multiple parameters and operating conditions.

\section{Conclusion}
This paper focuses on the development of a safe and robust GFL-GFM switching control strategy for GCIs under various operating conditions from the perspective of SSSR analysis. By considering the comprehensive circuit and control dynamics, a full-order small-signal state-space switched system model for GCIs under GFL-GFM switching control is established. Subsequently, based on this mathematical model, a characterization methodology for the SSSR and ISMD of the switched system is proposed, systematically revealing the impacts of multiple system parameters on the SSSRs and ISMDs of GCIs. Furthermore, a novel CSI integrating multiple system performance indicators, such as the stability margin, parameter sensitivity, and boundary distance, is formulated. This CSI facilitates the design of a multi-objective adaptive GFL-GFM switching control strategy to ensure the dynamic security and robustness of the GCI system. In the case study for a structured EMT simulation model of the GFL-GFM switchable inverter connected to an infinite bus on the CloudPSS platform, the effectiveness of the proposed GFL-GFM switched system model, the SSSR and ISMD analysis methods, and the CSI-based switching mechanism are verified.

\section*{Appendix}
The GCI system operates under the rated conditions. The detailed system parameters are listed in Table \ref{Table1_SSSR}.

\begin{table}[!t]
    \setstretch{0.91}
	\caption{Parameter Settings for GFL-GFM Switched System of GCI}
	\label{Table1_SSSR}
	\centering
	\resizebox{\columnwidth}{!}
	{\begin{threeparttable}
		\begin{tabular}{cccc}
			\toprule
			\toprule
			Type & Symbol & Description & Value (p.u.)\\
			\midrule
			\multirow{5}{*}{\makecell[c]{System\\Parameters}}
            & $R_f$
            & Filter resistance
            & 6.89$\times$10$^{-\text{4}}$ \\
            & $L_f$
            & Filter inductance
            & 0.54 \\
            & $C_f$
            & Filter capacitance
            & 0.067 \\
            & $SCR$
            & Short circuit ratio
            & 5.0 \\
            & $X/R$
            & Reactance resistance ratio
            & 5.0 \\
            
            \midrule
			\multirow{6}{*}{\makecell[c]{GFL\\Parameters}}
            & $K_{p\rm{PLL}}$
            & Proportional gain of PLL
            & 0.5 \\
            & $K_{i\rm{PLL}}$
            & Integral gain of PLL
            & 1$/\pi$ \\
            & $K_{po1}$
            & Proportional gain of OPL
            & 0.01 \\
            & $K_{io1}$
            & Integral gain of OPL
            & 1$/\pi$ \\
            & $K_{pi1}$
            & Proportional gain of ICL
            & 1.0 \\
            & $K_{ii1}$
            & Integral gain of ICL
            & 10$/\pi$ \\
            
            \midrule
			\multirow{6}{*}{\makecell[c]{GFM\\Parameters}}
            & $J$
            & Virtual moment of inertia
            & 1$/$100$\pi$ \\
            & $K_D$
            & Virtual damping coefficient
            & 20 \\
            & $K_{po2}$
            & Proportional gain of OVL
            & 1.0 \\
            & $K_{io2}$
            & Integral gain of OVL
            & 1$/\pi$ \\
            & $K_{pi2}$
            & Proportional gain of ICL
            & 10.0 \\
            & $K_{ii2}$
            & Integral gain of ICL
            & 1$/\pi$ \\
			\bottomrule
			\bottomrule
		\end{tabular}
	\end{threeparttable}}    
    \vspace{-2mm}
\end{table}

\bibliographystyle{IEEEtran}
\bibliography{References}

\end{document}